\documentclass[article]{JHEP3}

\newcommand{\Yh}{\ensuremath{\hat{Y}}}

\newcommand{\eref}[1]{(\ref{#1})}

\title{Quantum Mechanics of Yano tensors: Dirac equation in curved spacetime}  

\author{Marco Cariglia 
        \\ 
        DAMTP, Centre for Mathematical Sciences, Cambridge University
        \\ Wilberforce Road, Cambridge CB3 OWA, UK \\

  E-mail: \email{M.Cariglia@damtp.cam.ac.uk}}

\preprint{DAMTP-2003-146}

\abstract{In spacetimes admitting Yano tensors the classical theory of the spinning 
particle possesses enhanced worldline supersymmetry. Quantum mechanically  generators of 
extra supersymmetries correspond to operators that in the classical limit commute with 
the Dirac operator and generate conserved quantities. We show that the result is preserved 
in the full quantum theory, that is, Yano symmetries are not anomalous. This was known 
for Yano tensors of rank two, but our main result is to show that it extends to Yano tensors 
of arbitrary rank. We also describe the 
conformal Yano equation and show that is invariant under Hodge duality. There is a natural 
relationship between Yano tensors and supergravity theories. As the 
simplest possible example, we show that when the spacetime admits a Killing spinor then this 
generates Yano and conformal Yano tensors. As an application, we construct Yano tensors 
on maximally symmetric spaces: they are spanned by tensor products of Killing vectors.}


\begin{document}
 \section{Introduction.}

\noindent The concept of symmetry, and especially isometry, has
been extensively used in General Relativity in the
past. However, already in 1979 Collins \cite{Collins} pointed
out  that the ordinary use of symmetries had reached a ``stagnation
point'', and 
few further results could be obtained along the same direction. 
Symmetries of a Hamiltonian system in curved spacetimes are in one to one 
correspondence to 
Killing tensors. These are symmetric tensors $K^{\mu_1 \dots \mu_n}=
K^{(\mu_1\dots \mu_n)}$, associated to conserved quantities of degree $n$ in 
the momentum variables, and generate symplectic transformations in the phase 
space of the system. These transformations are determined by a vector flow. 
Ordinary Killing vectors generate isometries and correspond to the pushforward 
on coordinate space of the vector flow. Higher rank Killing tensors instead, 
which are defined by the condition $D_{(\lambda }K_{\mu_1 \dots \mu_n )} =0$, 
have zero pushforward: they correspond to genuine symmetries of the
whole phase space, and not only of configuration space. 
 
\noindent The next simple object that can be studied on a manifold $M$
after Killing tensors is a Yano
tensor. Such objects were introduced from a purely mathematical
point of view in 1951 by Yano \cite{Yano}. The physical interpretation
has remained obscure until Floyd \cite{Floyd} and Penrose
\cite{Penrose} showed that the Stackel-Killing tensor $K^{\mu\nu}$ of the
four-dimensional Kerr-Newman spacetime admits a square root, that is an
antisymmetric tensor $f^{\mu\nu}=f^{[\mu\nu]}$ such that 
	\begin{equation}
	K^{\mu\nu}=f^{\mu\lambda}\, f^{\nu}_{\;\; \lambda} . 
	\label{eq:f_squared} 
	\end{equation}
$f^{\mu\nu}$ proves indeed to be a Yano tensor and Carter and
McLenaghan \cite{CarterMcLenaghan} were then able to
construct from it a linear differential operator that commutes with
the Dirac operator. This provides a further quantum number for the
spinor wavefunction and explains why separation of variables can be
achieved for the Dirac equation on that background.
Gibbons, Rietdijk and van Holten \cite{GibbonsRietdijkVanHolten} gave
a systematic description of the role of rank 2 Yano tensors in General
Relativity. They
generate additional supercharges in the dynamics of
pseudo-classical spinning particles moving in $M$. The
Yano  condition
ensures that these supercharges are superinvariant. They generate
a classical superalgebra (in the sense of Poisson-Dirac brackets) that
closes on conserved quantities associated to
Killing tensors of rank $2$. Tanimoto \cite{Tanimoto} generalized the
construction to Yano tensors of generic rank $p$, which always
generate rank $2$ Killing tensors when squared. Yano tensors generated
by the action of groups on a spacetime have been discussed for example
in \cite{Macfarlane_Azcarraga}, \cite{Macfarlane}.

Two ingredients of the construction outlined above are physically
appealing and make the further study of Yano tensors a worthwhile
task.

\noindent The first is the connection between Yano tensors and
supersymmetry, both at the classical and quantum level. Yano
tensors are paired to their corresponding
Killing tensors (hence in some loose sense ``superpartners'') , in the
same way that the Dirac operator is paired to
its square, the Hamiltonian $H$. They also allow us to consider at the same
level the equation of motion for scalar and spinor wave functions: on one 
side there is the Carter-McLenaghan like operator which commutes with the 
Dirac operator and hence with the spinorial Hamiltonian. On the other side, 
from the square of the Yano tensor one can construct an operator acting on 
scalars of the form $K= K^{\mu\nu}\, D_\mu \, D_\nu$, where $K^{\mu\nu}$ is 
given by \eref{eq:f_squared} and $D_\mu$ is the covariant derivative on $M$. 
Then $K$ automatically commutes with the scalar Laplacian $\Delta = D_\mu 
g^{\mu\nu} D_{\nu}$. This is quite unexpected since in general the operator 
$K$ does not define a genuine quantum mechanical symmetry. The reason is 
that on a generic curved spacetime $M$ there appears a quantum anomaly 
proportional to a contraction of $K^{\mu\nu}$ with the Ricci tensor. 
However, when the Killing tensor $K^{\mu\nu}$ is of the form 
\eref{eq:f_squared}, then the anomaly disappears thanks to an integrability 
condition satisfied by the Yano tensor. 
Most important of all, Yano tensors generate an  {\it exotic
superalgebra}, which is more general than an usual extended
superalgebra. 
 
The second ingredient is the fact that Yano tensors, with their non
standard superalgebra, allow us to consider extended supersymmetry in 
manifolds
more general than K\"{a}hler. Usually one considers String Theory,
which  can be thought of as a
two-dimensional supersymmetric sigma model, where demanding to have
extended supersymmetry on the worldsheet
always leads to a K\"{a}hler structure in target space. Instead for the 
spinning particle, which is a one dimensional supersymmetric sigma
model, having a smaller worldsheet implies a greater set of
possible target spacetimes, that are all those which admit Yano tensors.
Therefore, Yano tensors are of direct interest in the study of
manifolds that are less restricted than K\"{a}hler ones. 

In this paper we begin a systematic study of Yano tensors and their relation
with Quantum Mechanics. 
\noindent 
The results quoted so far about Yano tensors are purely classical,
that is the supercharges associated to them generate extra
supersymmetries at the level of Poisson-Dirac brackets. Only in the
case of rank $2$ Yano tensors the result of Carter and
McLenaghan is quantum mechanical. In principle, when quantizing the theory
of the spinning particle, quantum mechanical anomalies  could appear
in the commutation relations due to the fact that spacetime is not flat. 
The first step in understanding the role of Yano tensors then
becomes that of analysing what happens at the quantum mechanical
level. 
What we discover and present here as the main result obtained is that
indeed anomalies are
completely absent, {\it
independently} of the rank of the Yano tensor considered. The result
is quite remarkable since in principle there is no particular reason
why the conservation laws should be quantum mechanically protected. 

\noindent  
There seems to be a relationship between the absence of anomalies and
the fact that Yano tensors are completely antisymmetric objects. 
Ordinary conservation laws associated to symmetric tensors, like the
energy momentum tensor or Killing tensors, are usually plagued by
quantum mechanical anomalies. What emerges here instead is that for
antisymmetric objects, which are somehow more fundamental, there are
no anomalies. Nevertheless, we still lack a clear geometrical
understanding of why this happens. 

\noindent   
The main features of the system remind the dynamics of $p$--forms on a
compact Riemannian manifold $V$. The bundle of differential forms
$\Lambda^* (V)$ over $V$ can be
decomposed  into the bundle of even and odd forms. Then the operator
$d+\delta $, where $d$ is the exterior derivative and $\delta$ its
adjoint, is elliptic and defines the de Rham complex on
$V$. Supersymmetry can be implemented by defining a real (Majorana)
supercharge to
be $Q=d+\delta :\Lambda^{even} (V)\rightarrow \Lambda^{odd} (V)$. Then
$Q^\dag = d+\delta :\Lambda^{odd} (V)\rightarrow \Lambda^{even} (V)$
and the superalgebra is simply
$\left\{ Q, Q^\dag \right\} = 2\,\Delta := 2\, H$. Energy is bounded from
below by zero. 'Bosonic'
and 'fermionic' degrees of freedom at the same at energy level $E\neq
0$ are paired, since $Q$ is an elliptic operator. However, in order to
see if supersymmetry is preserved one has to consider the Witten
index, which coincides with the index of the Laplacian.  It is well
known that the statement about the index is purely
topological, it does not depend on the metric on $V$, and the reason is
that one is dealing with antisymmetric
objects, the $p$-forms. Therefore, once an appropriate manifold $V$ is
chosen, supersymmetry will be present with {\it no anomalies},
independently of the metric one puts on it. This is directly analogous
to what happens with Yano operators: in that case, the only possible
anomalies that could appear are proportional to the Riemann
tensor. Their absence can in principle mean that the underlying
structure is metric independent.

\noindent It then seems evident that Yano operators play an important
role in quantum mechanics. They provide exactly conserved quantum
numbers for the Dirac equation and might be relevant in the study of
the scalar Laplace operator on $M$ and of its higher spin (both
integer and semi integer) counterparts, due to a profound geometrical
meaning which still needs to be uncovered. 
One possible geometrical setting where Yano tensors and their related
operators can acquire a proper meaning is that of complexes and index
theorems. Yano operators can be considered as exotic generalizations
of the Dirac operator $D\hspace{-0.22cm} /$. In some cases their square
defines a (pseudo) elliptic operator with an associated
index. Therefore they generate a new kind of complex which is very
special, since they commute/anticommute with the Dirac operator. In 
\cite{Kasper1} it has been shown that as a consequence Yano operators 
and Dirac operators have the same index. In \cite{Kasper2} one example 
of this correspondence, involving index theorems with torsion on Taub-NUT 
spaces, has been explicitly worked out.

Motivated by these considerations, we look for spacetimes admitting
Yano tensors. It is easy to show that this is true for all spacetimes of
almost special holonomy. Such spacetimes are of interest both from the
mathematical point of view, because
they are a rather general class of manifolds, and from the physical
point of view, since
they can arise as solutions for compactifications of supergravity
theories in
higher dimensions that preserve some supersymmetry in presence of
fluxes. For example one could use the new conserved quantities to
solve Dirac equation on singular (almost) $G_2$ spacetimes, and look
for chiral fermions.    
Main feature is that almost special holonomy spaces admit Killing
spinors. Given a Killing spinor, it generates a tower of antisymmetric
tensors of rank $1\leq p \leq D=dim(M)$,  which we show to be either
Yano or conformal Yano, depending whether $p$ is even or
odd. Conformal Yano tensors are
a generalization of Yano tensors. The set of conformal Yano tensors is
invariant under the Hodge duality operation and Yano tensors are in
correspondence with closed conformal Yano tensors. 
 
As a concrete example we realize this construction on maximally
symmetric spaces, all those that locally are given by spheres,
hyperbolic spaces, de Sitter and Anti-de Sitter. On such spaces it is
easy to solve the Killing equation explicitly and in a manifestly
coordinate independent way. Hence we show that Yano and conformal Yano
tensors on these spaces are extremely simple in form and given by a sum of
products of Killing vectors.

The rest of the paper is organized as follows. A substantial part of it is dedicated 
to a  review of background material of various kind. 
We begin with an extensive review of 
the theory of the classical spinning particle and its quantization in 
\ref{section:spinning_part}. The reader who is already acquainted with it can skip 
it altogether and read only sec.\ref{sec:Yano_op}, where quantum Yano operators are 
introduced. There we discuss the absence of anomalies. 
In \ref{sec:conformal_Yano} we derive the conformal Yano
equation and describe its relationship with the usual Yano equation
and Hodge duality. This turns out to be useful in the following
\ref{sec:almost_special_hol}, where we review some basic notions and then show that spaces 
of almost
special holonomy host a tower of antisymmetric tensors which are
either Yano tensors or conformal Yano. As a concrete application, we
explicitly construct such tensors on maximally symmetric spacetimes. 
In \ref{sec:maxim_symm} we present well known results about the geometry of symmetric
spaces and apply them in \ref{sec:special_tensors} to the solution of
the Killing equation when there is maximal symmetry. From Killing
spinors we construct Yano tensors and show that on these spacetimes
they are always given by a sum of tensor products of Killing vectors.      
We end in section
\ref{section:conclusions} with a summary and concluding remarks.

\section{The spinning particle \label{section:spinning_part}}
Here we describe the theory of the classical spinning particle in
curved spacetime and its quantization. We show that Yano operators
commute with the Dirac operator and therefore the quantum theory admits
genuine extra supersymmetries.

\subsection{The classical theory \label{section:classical} }
\noindent 
In this section we briefly summarize the classical theory of a spinning
particle in curved spacetime. Even if it is well known, it is the
natural setting where Yano tensors arise and acquire their physical
meaning. In the following, greek indices $\mu ,\nu ,\dots$ represent tensor
components in 
arbitrary reference frames while latin ones $a,b,\dots$ are referred
to locally Lorentzian frames. Conversion between the two is given by
using the vielbein $e_\mu^a (x)$.

Let $M$ be a spacetime. The spinning particle can be obtained as a
supersymmetric formulation of the ordinary point particle in $M$, and
is known to be the pseudo-classical limit of Dirac's theory of spin
$1/2$ fermions
\cite{BerezinMarinov,Casalbuoni,BarducciCasalbuoniLusanna,BrinkDeserZuminoDiVecchiaHowe,BrinkDiVecchiaHowe,vanHolten}.

The configuration space of the theory is given by a bundle which is
a fermionic extension of the tangent bundle $T(M)$. Fermionic
coordinates are special in that they behave at the same time as
coordinates and momenta, as is well known. Let 
$x^{\mu}$, $\mu =1, \dots ,D$, where $D=dim(M)$, be bosonic
coordinates for $M$. First of all one has to define a fermionic
extension of $M$ itself. Take $M_\psi$ to be the bundle with base $M$,
described
locally by the coordinates $(x^\mu , \psi^\nu )$, where the $\psi^\nu$ are
Grassmannian variables. Under a change of variables in the base
manifold $M$ the Grassmannian coordinates $\psi^\nu$ have the same
transition functions as a vector in the tangent space of $M$, $T(M)$,
and the bundle is well defined. $M_\psi$ is the configuration
space. A particle with spin moving in the
spacetime is described by a curve $\tau\mapsto (x(\tau),\psi(\tau))
\in M_\psi $, $\tau \in \mathbb{R}$. The Grassmann vector has the 
physical interpretation of being the {\it spin} of a particle, more
on this will be found below in this section. One needs to define a
connection on $M_\psi$, and the most natural choice is to use the same
connection of $T(M)$: 
\begin{equation} 
	\nabla_\mu \psi^\nu = \partial_\mu \, \psi^\nu +
	\Gamma_{\mu\rho}^\nu \, \psi^\rho . 
\end{equation} 
The curve describing the particle on $M_\psi$ can be obtained by a lift of 
the curve 
$\tau\mapsto x(\tau)$ by asking that $\psi^\nu$ is parallely transported: 
\begin{equation} 
	\frac{D \psi^\nu}{Dt} := \dot{x}^\mu \, \nabla_\mu \psi^\nu = 0 . 
	\label{eq:parallel_transp}
\end{equation} 
Geometrically this corresponds to splitting the tangent bundle of $M_\psi$, 
$T(M_\psi )$, into a {\it vertical} subbundle $T^v (M_\psi )$, with
fibre generated  by the 
vectors $\left\{ \partial /\partial \psi^\nu \right\}$, and a 
{\it horizontal} subbundle $T^h (M_\psi )$, with fibre generated by the 
vectors 
\begin{equation} 
	D_\mu := \partial_\mu - \Gamma_{\mu\nu}^\lambda \, \psi^\nu \, 
	\frac{\partial}{\partial \psi^\lambda} .
\end{equation} 
Then, along the lifted curve, for every function $f=f(x,\psi )$ on $M_\psi$ 
one has 
\begin{equation} 
	\frac{d f}{dt} = \dot{x}^\mu \, \frac{\partial f}{\partial x^\mu} + 
	\dot{\psi}^\nu \, \frac{\partial f}{\partial \psi^\nu} = 
	\dot{x}^\mu \, D_\mu (f) . 
\end{equation} 
This means, as is well known, that $T^h (M_\psi )$ is isomorphic to $T(M)$. 
Now it becomes easy to describe the lift  $\tau\mapsto (x(\tau))$ to the whole fermionic 
tangent bundle $T (M_\psi )$: first lifting it to $M_\psi $ using 
eq.\eref{eq:parallel_transp}, and then one choosing a connection on 
$T^h (M_\psi )$, defined by 
\begin{equation} 
	\nabla^h_\mu v^\nu:= \partial_\mu v^\nu + \Gamma_{\mu\lambda}^\nu \, 
	v^\lambda  .  
	\label{eq:h_connection}
\end{equation} 
We then ask for the lift of the curve to satisfy the 
equation 
\begin{equation} 
	\frac{D^2 x^\mu}{D\tau^2} := \dot{x}^\lambda \, \nabla^h_\lambda \, 
	\dot{x}^\mu = \frac{1}{2} \, R^\mu_{\;\;\nu} \, \dot{x}^\nu  , 
	\label{eq:parallel_transp2}
\end{equation} 
where we have defined $R_{\mu\nu}:=i/2 R_{\mu\nu \lambda_1 \lambda_2} 
\psi^{\lambda_1} \psi^{\lambda_2}$. 
Eqs.\eref{eq:parallel_transp}, \eref{eq:parallel_transp2} define generalized geodesics 
on $M_\psi$. The term $R^\mu_{\;\;\nu} \, \dot{x}^\nu$ comes from physical requirements, 
basically from the supersymmetry of the Lagrangian.  In fact, 
as in the bosonic case, there is a Lagrangian formulation that generates 
such curves. The Lagrangian density is nothing else than a supersymmetric 
extension of the Lagrangian density that gives geodesics on $M$. 
It is invariant under super-reparametrizations of $\tau$. A suitable gauge
choice gives the following form for the Lagrangian density:  
\begin{equation}
	L = \frac{1}{2} g_{\mu\nu}\dot{x}^\mu \dot{x}^\nu +
	\frac{1}{2}i g_{\mu\nu}\psi^\mu \frac{D\psi^\nu}{D\tau} . 
	\label{eq:lagrangian} 
\end{equation}
\noindent The gauge choice is expressed by two conditions: that the trajectory is null,  
and that spin is spacelike, 
\begin{equation} 
	\dot{x}^\mu \psi_\mu = 0.   \label{eq:constr2} 
\end{equation}   
The equations of motion that come from \eref{eq:lagrangian} are exactly 
\eref{eq:parallel_transp},
\eref{eq:parallel_transp2}. The particle is massless. No mass is
admitted since in the
original, super-reparametrization invariant Lagrangian it is not
possible to construct a fermionic mass term. However, at the quantum
level it will still be possible to impose a mass condition.  
 
In order to move to the Hamiltonian formulation and quantize the
system it is necessary to take into account that the momentum
conjugate to $\psi$ is proportional to $\psi$ itself. This is a second
class constraint that can be eliminated in a standard way and yields
the Poisson-Dirac brackets 
\begin{eqnarray} 
	\left\{x^\mu , p_\nu \right\} &=& \delta^\mu_\nu , \\ 
	\left\{\psi^\mu , \psi_\nu\right\} &=& -i\delta^\mu_\nu . 
	\label{eq:psi_bracket}
\end{eqnarray} 
However, $x^\mu$, $p_\nu$ and $\psi^\rho$ are not canonical coordinates since 
the bracket between $p$ and $\psi$ does not vanish. 
From now on we transform the Greek vector index of $\psi^\nu$ into a 
latin one using the vielbein, $\psi^a = e^a_\nu \, \psi^\nu$. This because 
we want to think of it as the spin of the particle, which is defined on 
locally Lorentzian frames. We quickly remark on the statement made at the 
beginning of the section, 
saying that $\psi^\nu$ represents the spin of the particle. Given the 
Poisson-Dirac brackets \eref{eq:psi_bracket}, these are (modulo constants) 
the same describing the Clifford algebra of Gamma matrices. Therefore, one 
readily sees that the operator $\psi^a \psi^b$ generates rotations on the 
spin vector via the Poisson-Dirac bracket, and that $R_{\mu\nu}$ as 
defined before generates transformations belonging to the holonomy group of 
the manifold.

In terms of the covariant momentum $\Pi_\mu = p_\mu -i/2 \omega_{\mu
ab}\psi^a \psi^b= \dot{x}_\mu$ (this formula becomes evident rewriting the 
Lagrangian in terms of $\psi^a$) it is possible to write the
Hamiltonian $H$ as 
\begin{equation} 
	H = \frac{1}{2} g^{\mu\nu} \Pi_\mu \Pi_\nu , 
\end{equation}
and the Poisson-Dirac bracket of two functions $F,G$ in the covariant
way 
\begin{equation} 
	\left\{F,G\right\}= \mathcal{D}_\mu F \frac{\partial G}{\partial\Pi_{\mu}} -
	\frac{\partial F}{\partial\Pi_{\mu}} \mathcal{D}_\mu G +
	R_{\mu\nu} \frac{\partial F}{\partial\Pi_{\mu}}  \frac{\partial
	G}{\partial\Pi_{\nu}} + i \, (-1)^{deg \, F}  \frac{\partial
	F}{\partial\psi^a}  \frac{\partial F}{\partial\psi_a} , 
	\label{eq:bracket} 
\end{equation} 
where the covariant derivative of a function is 
\begin{equation} 
	 \mathcal{D}_\mu F = \partial_\mu F +
	 \Gamma_{\mu\lambda}^{\rho}\Pi_\rho \frac{\partial F}{\partial
	 \Pi_\lambda} + \psi^a \omega_{\mu a}^{\;\;\;\; b}\frac{\partial F}{\partial
	 \psi^b} . 
\end{equation} 
From this and \eref{eq:bracket} it is seen that the phase space of the system is 
defined by a bundle which has 
$(\Pi_\mu , \psi_\nu )$ as coordinates for its fibers. 

\noindent The spin constraint \eref{eq:constr2} together with the fact
that the particle trajectory is null is compatible
with the equations of motion since such conditions are equivalent to
setting $H$
and $\mathcal{Q} = \Pi_\mu \psi^\mu$ equal to constants.

\subsection{Supersymmetry and Yano tensors} 
The Lagrangian \eref{eq:lagrangian} is invariant under supersymmetric
transformations generated by the Grassmann odd supercharge
$\mathcal{Q}$, which amount to 
\begin{eqnarray} 
	\delta x^\mu &=& i \epsilon  \left\{\mathcal{Q}, x^\mu\right\}= - i \epsilon
	\psi^\mu ,  \\ 
	\delta \psi^a &=& i \epsilon  \left\{\mathcal{Q}, \psi^a\right\}= \epsilon
	\dot{x}^a -  \delta x^\mu \omega_{\mu \,\,\, b}^{\, a} \psi^b . 
\end{eqnarray}   
It is also invariant under chiral symmetry, generated by 
\begin{equation} 
	\gamma_* = - \frac{ i^{[D/2]} }{D!}\, \eta_{a_1\dots
	a_D}\psi^{a_1}\dots \psi^{a_D} , 
\end{equation} 
where $\eta_{a_1\dots a_D}$ is the antisymmetric tensor (not density),
and dual supersymmetry, generated by 
\begin{equation} 
	\mathcal{Q}^* = i \left\{\mathcal{Q}, \gamma_*\right\} =
	- \frac{ i^{[D/2]}
	}{(D-1)!}\,  \eta_{a_1\dots a_D}\Pi^{a_1}\psi^{a_2}\dots
	\psi^{a_D} .  
\end{equation} 
$H$, $\mathcal{Q}$, $\mathcal{Q}^*$ and $\gamma_*$ provide the only
{\it generic} symmetries of the theory, i.e. those that exist for
every spacetime, see
\cite{vanHoltenRietdijk,vanHoltenRietdijk2}. 

Additional {\it
non generic} supersymmetries can be constructed in spacetimes which
admit Yano tensors. By definition a Yano tensor of rank $r$ is an
antisymmetric tensor 
\begin{equation} 
	f_{\mu_1\dots\mu_r} = f_{[\mu_1\dots\mu_r]} , 
	\label{eq:Yano_def1} 
\end{equation} 
such that the further condition 
\begin{equation} 
	\nabla_{( \mu_1} f_{\mu_2 )\mu_3 \dots \mu_{r+1} } = 0 , 
	\label{eq:Yano_def2} 
\end{equation} 
holds. This is equivalent to say that $f_{\lambda \mu_1\dots \mu_{r-1}
}\dot{x}^{\lambda}$ is parallely propagated along
geodesics\footnotemark .\footnotetext{In the introduction we argued
that the absence of quantum anomalies could be due to the fact there
exists an underlying structure that does not depend on the metric. As
can be seen from \eref{eq:Yano_def2}, Yano tensors {\it do} actually
depend on the metric. However, the Dirac operator depends on
the metric as well, and therefore the metric independent structure must
result by considering Dirac and Yano operators together.} 
Yano has studied these tensors and found conditions the
Riemann tensor has to satisfy in order for the spacetime to admit
any (\cite{Yano}). 
In \cite{GibbonsRietdijkVanHolten,Tanimoto} it has been shown that the
only additional non generic supercharges that are also supersymmetric
can be constructed from Yano tensors using the formula 
\begin{equation} 
	Y_r = \Pi^{\mu_1} f_{\mu_1
	\dots\mu_r}\psi^{\mu_2}\dots\psi^{\mu_r} -
	\frac{i}{r+1}\nabla_{\mu_1} f_{\mu_2 \dots\mu_{r+1}
	}\psi^{\mu_1}\dots \psi^{\mu_{r+1}} .  
	\label{eq:charges} 
\end{equation}
Analogously to Killing vectors (which are Yano tensors of rank $1$),
Yano tensors satisfy an integrability condition that comes from
eqs.\eref{eq:Yano_def1}, \eref{eq:Yano_def2}: 
\begin{equation} 
	\nabla_a \nabla_b f_{\mu_1\dots \mu_r} = (-1)^{r+1}
	\frac{r+1}{2} R^{\lambda}_{\;\; a [b \mu_1} f_{\mu_2\dots\mu_r]
	\lambda} . 
	\label{eq:int_cond} 
\end{equation} 
A typical example of rank $2$ Yano tensor is given in flat four dimensional 
Minkowski
space, where the associate conserved quantity has the interpretation
of an angular momentum. Let $v^{\mu}$ be a timelike vector associated
to the tangent of an observer's worldline, then the Yano tensor is  
\begin{equation} 
	f_{\mu\nu} = v^{\lambda} \,\eta_{\lambda\rho\mu\nu}\, x^{\rho} =
	\epsilon_{\rho\mu\nu}\, x^\rho   , 
	\label{eq:sol_flat_4d}
\end{equation}
where we have defined $\epsilon_{\rho\mu\nu}= v^{\lambda}
\,\eta_{\lambda\rho\mu\nu}$. Then the conserved quantity
$f_{\mu\nu}\,\Pi^\mu = \epsilon_{\rho\mu\nu}\, x^\rho \, \dot{x}^\nu$    
has the clear interpretation of an angular
momentum as seen by the observer. This is the reason why sometimes
quantities associated to 
rank $2$ Yano tensors are referred to as generalized angular momenta. 
 
\noindent The supercharges \eref{eq:charges} generate an exotic
extended superalgebra, in the sense that anticommutators of two
supercharges do not necessarily close on the Hamiltonian. The case of
$\mathcal{N} = q$ supersymmetry generated by rank two Yano tensors
$f_i^{\mu\nu}$, $i=1, \dots , q$ has been studied in
\cite{GibbonsRietdijkVanHolten}. The result is 
\begin{equation} 
	 \left\{\mathcal{Q}_i , \mathcal{Q}_j  \right\} = - 2\, i\,
	 Z_{ij} , 
\end{equation} 
where the conserved charge $Z_{ij}$ is given by 
\begin{eqnarray} 
	 Z_{ij} &=& \frac{1}{2} \, K_{ij}^{\mu\nu} \, \Pi^\mu \,
	 \Pi^{\nu} + (\mbox{spin corrections}) , \label{eq:Killing_charge}  \\ 
	K_{ij}^{\mu\nu} &=& \frac{1}{2} \, (f_{i \lambda}^\mu \,
	 f_{j}^{\nu\lambda} + f_{j \lambda}^\mu \,
	 f_{i}^{\nu\lambda}) .  \label{eq:f_squared2} 
\end{eqnarray} 
$K_{ij}^{\mu\nu}$ is a Killing vector, as it is easy to check using
\eref{eq:Yano_def1}, \eref{eq:Yano_def2}, and $Z_{ij}$ the associated
conserved quantity for a particle with spin. The vielbein $e^a_\mu$,
once transformed into an object with both indices curved (that is, the
metric), 
though not antisymmetric still defines via \eref{eq:charges} the 
conserved charge $\mathcal{Q}$, such that $\left\{\mathcal{Q}
, \mathcal{Q}   \right\} = - 2\, i \, H$. When one of these Killing tensors is different
from the metric tensor and invertible, then there exists a
dual spacetime where $K^{\mu\nu}$ is an (inverse) metric and
$g^{\mu\nu}$ a Killing tensor. Then it can be seen that the Yano
tensor associated to $K^{\mu\nu}$ becomes a vielbein in the dual
space, while the vielbein corresponds to a Yano tensor (see for
example \cite{Riet_VanHolt}, \cite{Baleanu}).  Not always the classical
symmetry associated to a general Killing tensor will survive at the
quantum level, as we will see
in the next section.

\subsection{The quantum system \label{section:quantum}} 
In this section we discuss the quantization of the spinning
particle. First we show how it implies the Dirac equation. Then in the 
following we
define quantum operators $\Yh_r$ which correspond to the classical
conserved charges 
of \eref{eq:charges}, and calculate the
(anti)commutator $\left[ D\hspace{-0.22cm}/ ,\Yh_r
\right. \left. \right\}$, where $D\hspace{-0.22cm}/  = \Gamma^\mu \,
d_\mu$ is the Dirac operator. The (anti)commutator is
identically  zero
for every $r$. We comment on the relevance of this result to the
theory of the spinning particle: the extra supersymmetries are actually
constraints on the Hilbert space of the system that reduce the number
of its degrees  of freedom.

\noindent As already mentioned above, the lagrangian \eref{eq:lagrangian} can be
obtained by a super reparametrization invariant lagrangian after a
gauge fixing (\cite{BrinkDiVecchiaHowe}) that amounts to a mass shell
condition 
\begin{equation} 
	2H = g^{\mu\nu}\Pi_\mu \Pi_\nu = 0 \;\;  , 
\end{equation} 
and a condition of transversality for the spin 
\begin{equation}
	Q= \psi^\mu \Pi_\mu = 0 . 
\end{equation}
This describes the massless case. In the quantum theory $H$ and $Q$
will become operators and we can relax the assumption by simply asking
that they are constant on the Hilbert space of the system: $Q=\pm m$,
$H=-m^2$. 
There are two ways to quantize the Poisson-Dirac brackets
\eref{eq:bracket}. As a specific example we consider the case where the
spacetime $M$ is four dimensional, but the same procedure apply in
generic dimension. 
The first type of quantization consists in choosing  
\begin{equation}
	\psi_\mu = \frac{1}{\sqrt{2} }\Gamma_\mu , 
	\label{eq:gamma} 
\end{equation}
where $\Gamma_\mu$ are the gamma matrices. In this case the algebra of
Grassmannian variables is substituted by the Clifford algebra, which
is richer. Basic (anti)commutators then are  
\begin{eqnarray} 
	\left[ \hat{q}_\mu , \hat{p}_\nu \right] &=& i\hbar\, g_{\mu\nu} , \\ 
	\left\{ \Gamma_\mu , \Gamma_\nu \right\} &=& 2\hbar\,  g_{\mu\nu}
	, \label{eq:anticomm} 
\end{eqnarray} 
where we have made explicit the factor $\hbar$ in order to keep track
of it along the following calculations. The covariant momentum operator 
$\Pi_\mu$ is
represented by the spinor covariant derivative $D_\mu = \partial_\mu +
\frac{1}{4} \omega_{\mu}^{\;\; ab}\Gamma_{ab}$. 
The constraints amount to $p^2=-m^2$ the first, while the second one
becomes 
\begin{equation}
  \Gamma^\mu D_\mu |\psi > = \pm m \, |\psi > ,   
\end{equation}
that is, this is the way the Dirac equation in spacetime arises in the
theory. The situation is similar to
that of the superstring where both spacetime fermions and bosons arise
as states of the system. As a matter of fact the spinning
particle as well has a bosonic sector which is quantized by taking 
\begin{equation}
	\psi_\mu = \frac{1}{2}\left( a_\mu + a_\mu^\dag \right) , 
	\label{eq:bosonic_quant} 
\end{equation}
for two operators $a,\, a^\dag$ that satisfy 
\begin{eqnarray} 
	\left\{ a_\mu, a_\nu^\dag \right\} &=& \hbar\, g_{\mu\nu} , \\ 
	\left\{ a_\mu, a_\nu \right\} &=& 0 = \left\{ a^\dag_\mu,
	a^\dag_\nu \right\} . 
\end{eqnarray} 
Then the gauge fixing amounts to select a physical subspace of the
full Hilbert space generated by vectors $|\psi >$ such that  
\begin{eqnarray}
  p^2 | \psi > &=& -m^2, \\ 
  p_\mu a^\mu |\psi > &=& \pm m \, |\psi >  ,   
  \label{eq:transv_cond}
\end{eqnarray} 
for example, for $m=0$ states are massless and transverse. 
Supersymmetry on the worldsheet does not necessarily imply
supersymmetry in target space. Take $m\neq 0$ for example. If
spacetime has $D$ dimensions the 
bosonic excitations are $p$--forms, $0\leq p \leq D$, and the total
number of degrees of freedom is given by $\Sigma_n {D-1 \choose n}  
 = 2^{D-1}$. Massive fermions
instead have $2^{[D/2]}$ degrees of freedom, or less if some Weyl or Majorana 
condition is used. The two numbers
coincide only for $D=1,2$. If $m=0$ instead fermionic degrees of
freedom are the same while bosonic ones are halved, and they coincide
for $D=3,4$. Hence, unless some kind of GSO projection is employed, 
in all the other cases there is no target space supersymmetry.

Lastly we note that, even if the conservation law $\left\{\mathcal{Q}
, H   \right\} = 0$ is satisfied for the quantum system, it is in
general not true that the operator associated to the classical charge
\eref{eq:Killing_charge} will commute with the Hamiltonian. For
example, even for a scalar particle this does not happen. One can
construct the operator $K= K^{\mu\nu}\, D_\mu \, D_\nu$, where $D_\mu$ is
the scalar covariant derivative. The Hamiltonian of the scalar
particle is given by $H= 1/2\, g^{\mu\nu}\, D_\mu \, D_\nu$, and
calculating  the commutator $\left[ K, H \right]$ one finds a quantum
anomaly proportional to 
\begin{equation} 
	\left[ K, H \right] \sim  K^\lambda_{[\mu}\, R_{\nu ] \lambda} , 
	\label{eq:anomaly}
\end{equation} 
where $R_{\mu\nu}$ is the Ricci tensor. Therefore in general if the
spacetime is curved the classical conservation law is violated quantum
mechanically. Operators constructed from symmetric tensors are in
general a source of anomaly. However, as noted in \cite{CarterMcLenaghan}, if
$K^{\mu\nu}$ is the square of a rank $2$ Yano tensor then the
contribution \eref{eq:anomaly} is identically zero due to the
integrability condition \eref{eq:int_cond}. This property is actually
true for a Yano tensor of arbitrary rank. In the next section we will
show that the classical supercharges \eref{eq:charges} are always
conserved even at the quantum mechanical level.

\subsection{Yano operators \label{sec:Yano_op}} 
When the spacetime admits some Yano tensors $f$ of rank
$r$, one can construct the quantities $Y_r$ as in
\eref{eq:charges}. They are defined in the symplectic space of the
system and classically commute both with the supercharge
$\mathcal{Q}$, which quantum mechanically corresponds to the Dirac
operator (eq.\eref{eq:gamma}), and its square the Hamiltonian. 
In quantizing the system to the $Y_r$'s there remain associated
operators $\Yh_r$ given by 
\begin{equation} 
	\Yh_r = (i)^{[\frac{r+1}{2}]} \left[
	\Gamma^{\mu_1\dots\mu_{r-1}}  f_{\mu_1
	\dots\mu_{r-1}\mu_r} \hbar D_{\mu_r}  -  
	\frac{(-1)^r}{2(r+1)} \Gamma^{\mu_1\dots\mu_{r+1}}
	\nabla_{\mu_1} f_{\mu_2 \dots\mu_{r+1}} \right] ,   
	\label{eq:Q_charges} 
\end{equation} 
where $D_\mu$ is the spinor covariant derivative and $\Gamma^{\mu_1\dots\mu_n}$ 
is the antisymmetrized product of $n$ Gamma matrices with unit weight. 
For $r$
even these operators contain an odd number of Gamma matrices and are
intrinsically 'fermionic', while for $r$ odd they are
'bosonic'. The factor of $i$ is chosen so that in
positive definite signature the operators are self-adjoint. In
Lorentzian signature there's always an anti self-adjoint part
proportional to $\Gamma^0$. 
The following lemma holds. 
 
\noindent {\bf Lemma.} Let $(M,g)$ be a manifold with spin structure
and let $f$ be a Yano tensor of rank $r\geq 1$, with associated
operator $\Yh_r$ as in \eref{eq:Q_charges}. Given the Dirac operator
$D\hspace{-0.22cm} / = \Gamma^\mu D_\mu$ then  
\[ \left[  D\hspace{-0.22cm} / \right. , \left. Y_r \right\} \equiv 0 ,
\] where $\left [ \; , \right. \left. \right\}$ is the commutator for $r$
odd and the anticommutator for $r$ even.

\noindent {\it Proof:} We delegate the proof to the appendix. $\triangle$ 
 
\vspace{\baselineskip} 
\noindent Therefore all the classical extra supersymmetries of the
theory of the 
spinning particle extend to quantum mechanical symmetries. The
implication of the lemma is even deeper in that it means that on
spacetimes admitting Yano tensors it is always possible to form
conserved charges for the Dirac equation. Geometrical properties of
spacetime and particle physics have so far always being
linked. Consider for example Kaluza Klein compactifications of
supergravity. Then the
isometries of the compact part of spacetime determine the number of massless
gauge bosons, while its holonomy group determines the number of
supersymmetries and of massless gravitinos. Now the next question is:
what is the geometrical meaning of the conserved quantities associated
to Yano tensors, and what is their role in particle physics? 
Yano tensors can exist on spaces which have no isometries at all. For
example six dimensional almost K\"{a}hler spaces admit  Yano tensors
of all ranks from $1$ to $5$. 
 
However, the situation is quite different from that of
ordinary quantum field \linebreak theory. While in quantum field
theory the supercharges carry
spacetime spinor indexes  and 
acting with them on states of the system corresponds to moving in a 
determined target space supermultiplet, in the spinning particle system 
the supercharges are target space scalars and do not change the spin
of a particle. They are associated to a split of each eigenspace of the 
Hamiltonian into a sum of two-dimensional eigenspaces on which the
supercharge, opportunely rescaled, acts as the group $\mathbb{Z}_2$. 
Then eigenstates of $Q$, which correspond to classical trajectories of 
the particle with a fixed conserved number, are given by linear
combinations of such states. 
 
 In the rest of the paper we start  analysing a class
of spacetimes which admit Yano tensors: spaces of almost special
holonomy. Among these, we will construct explicitly Yano tensors on
maximally symmetric spaces. On spaces of almost special holonomy one
can show that it is present a tower of antisymmetric tensors. These
will prove to be either Yano, or conformal Yano tensors. These latter 
objects satisfy a generalized Yano equation, the conformal Yano
equation. Yano tensors and conformal Yano tensors are related by
Hodge duality: as we show in the next section, the conformal Yano equation
is duality invariant.

\section{Conformal Yano equation and Hodge duality
\label{sec:conformal_Yano}} 
Generalizing the construction of eq.\eref{eq:f_squared2}, given two
rank $r$ Yano tensors $f_{\mu_1 \dots\mu_r}$, $g_{\mu_1
\dots\mu_r}$, it is possible to associate to them the symmetric rank $2$
tensor  
\begin{equation} 
	K^{(f,g)}_{\mu\nu} := ( f_{\mu\lambda_2 \dots\lambda_r}
	g_{\nu}^{\;\;\lambda_2 \dots\lambda_r} + g_{\mu\lambda_2 \dots\lambda_r}
	f_{\nu}^{\;\;\lambda_2 \dots\lambda_r} ) . 
	\label{eq:Killing_def} 
\end{equation} 
Eq.\eref{eq:Yano_def2} then ensures that $K^{(f,g)}$ is a Killing
tensor, i.e. it satisfies 
\begin{equation} 
	\nabla_{( \lambda}  K^{(f,g)}_{\mu\nu )} = 0 . 
\end{equation} 
It  is possible to use the same reasoning in order to construct
conformal Killing tensors. A conformal Killing tensor $K_{\mu\nu}$ is
by definition a symmetric tensor satisfying the equation 
\begin{equation} 
	\nabla_{( \lambda}  K_{\mu\nu )} = g_{( \lambda \mu }a_{\nu )} . 
\end{equation} 
Note that $K_{\mu\nu}+\alpha\, g_{\mu\nu}$ satisfies the same equation
for any given constant $\alpha$  and
so we can always take $K$ to be traceless. In this case then $a_\mu =
2/(D+2) \nabla^\lambda K_{\lambda\mu}$. As Killing tensors generate conserved charges
of the system, i.e. functions which commute with the Hamiltonian,
conformal Killing tensors generate functions whose commutator with $H$
is proportional to $H$ itself.  
The conformal Yano equation has been discussed by Tachibana
\cite{Tachibana}.  In the case of rank $2$ it is given by 
\begin{equation} 
	\nabla_{(\lambda} f_{\mu )\nu} = g_{\lambda\mu}\phi_\nu -
	g_{\nu (\lambda }\phi_{\mu )} , \label{eq:CY_def} 
\end{equation} 
where $\phi_\mu = 1/3
\nabla_{\lambda}f^{\lambda}_{\,\,\mu}$. Four dimensional spacetimes admitting a
conformal Yano tensor of rank $2$ must be of Petrov type D, N or O
(see for example \cite{GlassKress}). Moreover, eq.\eref{eq:CY_def} is
invariant in form under Hodge duality (\cite{FerrandoSaez}). 

Now we generalize the construction to conformal Yano tensors of higher
rank in arbitrary dimension. We can start by a definition similar to
that of eq.\eref{eq:CY_def}:  
\begin{equation} 
	\nabla_{( \mu_1} f_{\mu_2 )\mu_3 \dots \mu_{r+1} } = g_{\mu_1
	\mu_2} \Phi_{\mu_3 \dots \mu_{r+1} }+ A g_{[\mu_3  (\mu_1
	}\Phi_{\mu_2 )  \mu_4\dots \mu_{r+1} ]} ,  
	\label{eq:conformal_Yano_def} 
\end{equation} 
where $A$ is a constant and $\Phi = 1/(D+A) \nabla f$. This definition ensures that the tensor
$K^{(f,f)}$ as in eq.\eref{eq:Killing_def} satisfies the Killing
conformal equation with $a_\nu = 2 f_{\nu\lambda_1 \dots \lambda_{r-1}
}\Phi^{\lambda_1 \dots \lambda_{r-1} }$ independently of the constant
$A$. However, in general if $f,g$ are two rank $r$ antisymmetric
tensors and both satisfy
eq.\eref{eq:conformal_Yano_def}, then  $K^{(f,g)}$ is not conformal
Killing unless 
for both $f$ and $g$ the constant $A$ is equal to zero. 

In the appendix we show that the constant $A$ is fixed by consistency with Hodge
duality to be equal to $(1-r)$, and that, moreover, in this case
the dual of $f$ satisfies the same equation, that is, the conformal Yano
equation 
\begin{equation} 
	\nabla_{( \mu_1} f_{\mu_2 )\mu_3 \dots \mu_{r+1} } = g_{\mu_1
	\mu_2} \Phi_{\mu_3 \dots \mu_{r+1} } - (r-1) g_{[\mu_3  (\mu_1
	}\Phi_{\mu_2 )  \mu_4\dots \mu_{r+1} ]}   
	\label{eq:conformal_Yano_def2} 
\end{equation}
is preserved by duality. Yano tensors themselves are a subset of all
conformal  Yano tensors, those for which $\Phi = 0$ (co-closed). Whenever the
spacetime  admits a Yano tensor, then its dual is a conformal Yano
tensor.

\section{Spaces of almost special holonomy \label{sec:almost_special_hol}} 
Let $M$ be a spacetime of dimension $D$ which admits a spin structure
and $S$ a spin
bundle over $M$. If a Killing spinor exists on $S$ then its holonomy
is reduced with respect to a connection that extends the usual
Levi-Civita spinorial connection: the space is said to be of almost
special holonomy. In this situation  it is possible
to construct from it a tower of tensors on $M$, which happen to
split into genuinely Yano tensors and conformal Yano tensors. Hodge
duality mixes them along the lines of section
\eref{sec:conformal_Yano}. The local form of Yano and conformal Yano
tensors in pseudo-Riemannian spaces of constant curvature has been
studied in \cite{Stepanov}.    
 
We start from the definition of a Killing spinor, which is covariantly
constant with respect to the extended derivative 
\begin{equation} 
	\mathcal{D}^{\pm}_\mu \eta^\pm := (D_\mu \mp i\, \lambda
	\Gamma_\mu ) \eta^\pm = 0 , 
	\label{eq:Killing_eq} 
\end{equation} 
where $D_\mu$ is the spinorial covariant derivative. If $\lambda=0$
then this is the equation for a covariantly constant spinor, and all
the tensors one can construct from it are covariantly constant. 
 
It is worthwhile spending a few lines to comment on the integrability
conditions  of \eref{eq:Killing_eq} for two reasons. One is that in
this section we are going to construct special tensors on spacetimes
admitting Killing spinors and therefore we are interested in knowing
to which cases the construction applies. The second one is that in the
next section  we are going to solve the 
Killing equation for maximally
symmetric spacetimes, which result to be those that admit the maximum number of
solutions. Let's start with the case $\lambda = 0$,
that is, covariantly constant spinors. Then one has 
\begin{equation} 
	0 = \left[ D_\mu , D_\nu \right] \,\eta = \frac{1}{4}\,
	R_{\mu\nu\rho\sigma} \, \Gamma^{\rho\sigma} \,\eta . 
	\label{eq:der_comm} 
\end{equation} 
Multiplying on the left by $\Gamma^{\nu}$ one gets the condition 
\begin{equation} 
	R_{\mu\nu}\,\Gamma^\nu \,\eta = 0 , 
	\label{eq:condition_1} 
\end{equation} 
whose square gives  
\begin{eqnarray} 
	R_{\mu}^\lambda\, R_{\nu\lambda} &=& 0 , \label{eq:condition_2a} \\ 
	R &=& 0 .  \label{eq:condition_2b} 
\end{eqnarray} 
\eref{eq:condition_1}, \eref{eq:condition_2a}, \eref{eq:condition_2b}
give conditions  on the
Ricci tensor and the curvature scalar. There is also a condition on
the Weyl tensor that comes from eq.\eref{eq:der_comm}, amounting to 
\begin{equation} 
	\frac{1}{4}\,C_{\mu\nu\rho\sigma} \, \Gamma^{\rho\sigma}
	\,\eta = 0 . 
	\label{eq:condition_3} 
\end{equation}
In Euclidean signature \eref{eq:condition_2a} implies that spacetime
is Ricci flat and the holonomy of the spin connection $D_\mu$ is
reduced. Such spaces are well known. In particular, if $M$ is
simply-connected and its metric is irreducible and non symmetric,
there applies a classification due to Berger: depending on the number
of covariantly constant spinors and
the dimension of spacetime one can get the holonomy group to be
$SU(D/2)$ in even dimension (Calabi-Yau), $Sp(D/4)$ (hyperK\"{a}hler)
for $D=4m$ and the
exceptional cases $G_2$ for $D=7$, $Spin(7)$ for $D=8$(see
\cite{Joyce} for example).   
 
\noindent If the signature is Lorentzian instead
\eref{eq:condition_2a} does not imply that the Ricci tensor is
zero. A counterexample is given by the $pp$--wave spacetimes, whose
Ricci tensor is null. $pp$-wave spacetimes have $R=0$ and
eqs.\eref{eq:condition_1}, \eref{eq:condition_3}  become a null projection, so
that the number of covariantly constant spinors is at most half the maximum
number admitted. When the $pp$--wave spacetime is of Kowalski-Glikman
kind then it is a symmetric space (\cite{Ortin}). 
 
Now consider $\lambda \neq 0$. Conditions analogous to \eref{eq:condition_1}, 
\eref{eq:condition_2a}, \eref{eq:condition_3} are 
\begin{eqnarray} 
	R_{\mu\nu}\,\Gamma^\nu \,\eta &=& \left(
	2\,(D-2)\,\lambda^2+\frac{R}{2\, (D-1)} \right)   \,\Gamma_\mu
	\,\eta  \label{eq:condition_4a}     \\ 
	R_{\mu}^\lambda\, R_{\nu\lambda} &=&\left(
	2\,(D-2)\,\lambda^2+\frac{R}{2\, (D-1)} \right) \, R_{\mu\nu}
	, 	\label{eq:condition_4b}  \\ 
	\frac{1}{4}\, C_{\mu\nu\rho\sigma} \, \Gamma^{\rho\sigma} \,\eta   
	&=& 0 . 
	\label{eq:der_comm2}  
\end{eqnarray} 
These reduce to eqs.\eref{eq:condition_1}, \eref{eq:condition_2a},
\eref{eq:condition_2b}  by
taking $\lambda =0, R=0$. In Euclidean signature
\eref{eq:condition_4b} implies 
that the space is Einstein with constant scalar curvature $R= 4\, D(D-1)
\lambda^2$. All these spaces are locally symmetric and can be fully
described.

$\lambda$ real implies that $M$ is compact with positive
curvature, and can be classified by the holonomy of its cone, where the cone
over $M$ is defined as the space with metric $ds^2= dr^2+ r^2 \,
ds^2_M$. 
It is not
important whether the cone has any singularity at $r=0$ or not, since
one is simply interested in rewriting the Killing equation on $M$ into
a simpler equation on a different space. Therefore, all it is
necessary to do is to classify the possible cones admitting
covariantly constant spinors. The standard approach is to reduce this
procedure to Berger's classification: if one assumes $M$ is simply
connected (but this will not be true in general), then a result of
Gallot (\cite{Gallot}) says that the cone over $M$ is either flat, or
irreducible. A flat cone, outside the origin $r=0$, can be obtained
not only if $M$ is a $D$-dimensional round sphere, but for all spaces
that are locally a sphere, that is maximally symmetric
spaces. The sphere is the only space such that its cone has no
singularities at the origin, but maximal symmetry is a sufficient
condition for flatness outside $r=0$. If the cone is irreducible
instead, one can have $3$--Sasaki manifolds 
for $D=4m-1$ and hyperK\"{a}hler cone, Sasaki-Einstein for $D=4m\pm1$
and cone Calabi-Yau, almost K\"{a}hler for $D=6$ and cone $G_2$,
almost $G_2$ for $D=7$ and cone $Spin(7)$ (see for example
\cite{Figueroa}, \cite{Friedrich}).

For $\lambda$ imaginary  $M$ is non compact with
negative curvature. Then it can be shown that either $M$ is the
hyperbolic plane $H^D$ or it is given by a warped product $M=N\times
\mathbb{R}$ with metric $ds^2 = e^{\mu \, t}ds_N^2 + dt^2$, where $N$
is a complete, connected spin manifold which admits non trivial
parallel spinors and $\mu\in\mathbb{R}\setminus \left\{ 0 \right\}
$. Conversely, every such $N$ gives, when warped, a manifold $M$ with
imaginary Killing spinors (\cite{Friedrich}).

In Lorentzian signature instead the spacetime is not necessarily
Einstein and there is much more freedom. If we restrict our attention
to the subclass of Einstein spaces then it is possible again to relate
$\lambda$ and $R$. For $\lambda$ real the curvature is positive, even
if no compactedness is required, and for example the analogue of the
Euclidean sphere is given by de Sitter space. 
For $\lambda$ imaginary the curvature is negative and the analogue of
the hyperbolic plane is Anti-de Sitter space. The case $\lambda$
imaginary has been recently studied in \cite{Leitner}.

Because of the integrabililty conditions we just discussed, in general
only a subset of all the possible spinors on $M$ will satisfy
\eref{eq:Killing_eq}. For example, if there exists a solution of the
kind $\eta^+$, there is no reason in general to have also a solution
of the kind $\eta^-$, and vice versa (for maximally symmetric spaces
both exist). The subset of Killing spinors
will be selected by using an appropriate projection matrix. This implies
that, of all the tensor we are going to construct below, only part of
them will be non zero. However, to keep the maximum generality, we
will discuss all the possible cases.

We are now in the position to construct special tensors on spaces of
almost special holonomy. 
If the dimension $D=2m$ is even then one can construct a chirality
matrix 
\begin{eqnarray}
	\Gamma_* &=& (-i)^{m+t} \, \Gamma_1 \dots \Gamma_D , \\ 
	\Gamma_* \, \Gamma_* &=& 1 . 
\end{eqnarray} 
$t$ is zero in Euclidean signature and one in Lorentzian, which we
define using an almost plus convention. 
The chirality matrix satisfies 
\begin{equation} 
	\Gamma_* \, \mathcal{D}^{\pm} =\mathcal{D}^{\mp} \, \Gamma_* , 
\end{equation} 
and therefore the two kinds of Killing spinors can be related by 
\begin{equation} 
	\eta^- = \Gamma_* \eta^+ , 
\end{equation} 
and they are not eigenspinors of chirality. 
 
Regardless of the dimension and the signature one can define the
following $n$--forms 
\begin{eqnarray} 
	A^{(n)}_{[\mu_1\dots\mu_n ]}&:=& \overline{\eta}^+ \,
	\Gamma_{\mu_1\dots\mu_n} \eta^- , \label{eq:A_tensor}    \\ 
	B^{(n)}_{[\mu_1\dots\mu_n ]} &:=& \overline{\eta}^+ \,
	\Gamma_{\mu_1\dots\mu_n} \eta^+ \label{eq:B_tensor}  .  
\end{eqnarray} 
We want to show that for each $n=1,\dots , D-1$ either $A^{(n)}$ or
$B^{(n)}$ is a  Yano tensor and the other one is a conformal
 Yano. Moreover, if $A^{(n)}$ is Yano then $A^{(n+1)}$ is conformal
Yano, $A^{(n+2)}$ Yano again and so on, and the same for $B^{(n)}$. 

In the case of covariantly constant spinors these forms are known to
generate calibrations in supergravity theories without $p$--form
flux. In the presence of flux the same forms can be used to define
calibrations, but the spinors $\eta$ have to satisfy a generalized
Killing equation with flux, which we do not contemplate here. It
should be interesting to investigate further the connection between
Yano tensors and supergravity calibrations.

Note that condition $D_\mu \eta^+ = i \, \lambda \Gamma_\mu \eta^+$
implies 
\begin{equation} 
	D_\mu \overline{\eta}^+ = (-1)^{t+1} i\,
	\lambda^* \overline{\eta}^+  \Gamma_\mu  , 
	\label{eq:D_bar} 
\end{equation} 
so that in the following
discussion the tensors $A$ and $B$ will reverse their role depending
whether $t=0,1$ and whether $\lambda$ is real or imaginary. Moreover,
in case $\eta^\pm$ spinors are more than one
we will  distinguish them by an index $\alpha$, and \eref{eq:D_bar}
implies that it is
possible to normalize them. In Euclidean signature, and for 
$\lambda$ real (the $D$-dimensional sphere for example), one can take 
\begin{equation} 
	\overline{\eta}^{\alpha +} \, \eta^+_\beta = M^\alpha_\beta , 
	\label{eq:normalization} 
\end{equation} 
where $M^\alpha_\beta$ is a suitable constant matrix. For $\lambda$
imaginary (hyperbolic plane for example) one has to take the opposite
combination $\overline{\eta}^{\alpha +} \, \eta^-_\beta =
M^\alpha_\beta$, while passing from Euclidean to Lorentzian signature
these two cases interchange. 
 
For concreteness from now until the end of this section we now work in
Euclidean signature and with $\lambda$ real. To the extent of showing
properties of $A$ and $B$ tensors we use the
following identities  
\begin{eqnarray} 
	\Gamma_{\lambda}\, \Gamma_{\mu_1\dots\mu_n} &=&
	\Gamma_{\lambda\mu_1\dots\mu_n} + n g_{\lambda\, [\mu_1}\,
	\Gamma_{\mu_2\dots\mu_n]} , \label{eq:identities1} \\ 
	 \Gamma_{\mu_1\dots\mu_n} \, \Gamma_{\lambda} &=&
	\Gamma_{\mu_1\dots\mu_n\lambda} + (-1)^{n+1} n g_{\lambda\, [\mu_1}\,
	\Gamma_{\mu_2\dots\mu_n]} , 
	\label{eq:identities2} 
\end{eqnarray} 
together with \eref{eq:D_bar}, to calculate
covariant derivatives  of $A$ and $B$. 
Suppose $n=2k+1$ is odd. Then one has  
\begin{equation} 
	\nabla_{\lambda} A^{(n)}_{\mu_1\dots\mu_n} = - i\,\lambda
	\overline{\eta}^+ \left( \Gamma_{\lambda}\,
	\Gamma_{\mu_1\dots\mu_n} + \Gamma_{\mu_1\dots\mu_n} \,
	\Gamma_{\lambda} \right) \eta^-  = - 2\, i\, n\,\lambda
	\overline{\eta}^+ g_{\lambda\, [\mu_1}\,
	\Gamma_{\mu_2\dots\mu_n]} \eta^- ,  
	\label{eq:D_A} 
\end{equation} 
from which it is readily seen that 
\begin{equation} 
	\nabla_{( \lambda} A^{(n)}_{\mu_1 ) \mu_2 \dots\mu_n} = 
	 - 2\, i\, n\,\lambda \left[ g_{\lambda \mu_1} \overline{\eta}^+ 
	\Gamma_{\mu_2\dots\mu_n} \eta^- - (n-1)\, g_{[ \mu_2 \,
	 (\lambda} \overline{\eta}^+ \Gamma_{\mu_1 ) \, \mu_3\dots
	 \mu_n]} \eta^- \right] ,   
\end{equation} 
that is $A^{(2k+1)}$ satisfies the conformal Yano equation and is {\it not}
a strictly Yano tensor. However, from \eref{eq:D_A} it is immediate
to check that $A^{(2k+1)}$ is closed and therefore its Hodge dual is
strictly Yano. The same calculation for $B^{(2k+1)}$ yields 
\begin{eqnarray} 
	\nabla_{\lambda} B^{(n)}_{\mu_1\dots\mu_n} &=& i\,\lambda
	\overline{\eta}^+ \left( - \Gamma_{\lambda}\,
	\Gamma_{\mu_1\dots\mu_n} + \Gamma_{\mu_1\dots\mu_n} \,
	\Gamma_{\lambda} \right) \eta^+  = - 2\, i\, n\,\lambda
	\overline{\eta}^+ \Gamma_{\lambda\mu_1\dots\mu_n} \eta^+ \\ 
	&=& - 2\, i \lambda\, B^{n+1}_{\lambda\mu_1\dots\mu_n}    
	\label{eq:D_B} . 
\end{eqnarray} 
Therefore $B^{(2k+1)}$ is strictly Yano but, since it is not closed in
general,
its Hodge dual will be conformal Yano. 
For $n= 2k$ even instead the identities
eq.\eref{eq:identities1},\eref{eq:identities2}  imply
that $A$ and $B$ exchange their roles: therefore $A^{(2k)}$ becomes
Yano and $B^{(2k)}$ conformal Yano. Therefore one gets two towers
of tensors of increasing rank which are alternatively Yano and
conformal Yano. Yano tensors which are covariantly constant are 
more special than the others since they generate symmetries that
extend to String Theory.  It is not easy to 
give a general rule for the tensors of A and B kind to be covariantly 
constant. Clearly it is true when the Killing spinor is 
itself covariantly constant, or it can happen when the Killing equation 
gets modified by a flux term, like for pp-wave spacetimes. 
In absence of further information however a general spinor will not
necessarily give covariantly constant Yano tensors. One should take
into account other projections on the spinor, if they occur. 

It is conventional to call a spinor satisfying
\eref{eq:Killing_eq} Killing, since it generates Killing
vectors. However, Killing vectors happen to coincide with Yano tensors
of rank $1$ and, given that the whole tower of tensors constructed
from such a spinor is made of (conformal) Yano tensors, a more
appropriate name would rather be that of Yano spinor. 

\noindent Notice that in this setting we are dealing
with constructing conformal Yano tensors as covariant derivatives of Yano
tensors. This is not allowed in the general case because of the
integrability condition \eref{eq:int_cond}, but is possible on
spacetimes with Killing spinors.

One further remark is that $A$ and $B$ tensors have a
complementary behaviour which
is required by consistency with Hodge duality when the dimension is
even. When $D=2m$ the Hodge duality operation sends tensors of odd rank
into tensors of odd rank, and likewise for even rank, but at the same
time it exchanges tensors of kind $A$ and $B$, since it holds 
\begin{equation} 
	\Gamma_{\mu_1\dots\mu_n} = (i)^{m+t}\,\Gamma_*
	\frac{1}{(D-n)!} \eta_{\mu_1\dots\mu_n \nu_{n+1}\dots \nu_D}
	\Gamma^{\nu_D\dots\nu_{n+1}} .  
\end{equation} 
Therefore type $A$ tensors are sent into type $B$ tensors by duality,
but we know that duality changes Yano tensors into closed conformal
Yano tensors and vice versa. 
Since covariant derivatives of $A$ and $B$ depend on the algebra of
Gamma matrices, and this does not depend on the dimension being even
or odd, then one recovers the same complementary structure in odd
dimension.

\section{Maximally symmetric spaces} 
Maximally symmetric spaces are the simplest case to consider since
the absence of extra structures  makes the computation of
Killing spinors and (conformal) Yano tensors easy. Such spacetimes are
exactly those that support (when they are spin manifolds) the maximum
number of solutions of the Killing spinor equation. Using group
theoretical techniques and Gamma matrix manipulations we give a
sufficient condition for a symmetric space to be maximally symmetric,
in terms of Killing forms and structure coefficients.

In the following section we
give a brief account of some properties of symmetric spaces in general
and then show
how this can be used to solve the Killing spinor equation when the
space is maximally symmetric(see
\cite{Ortin} for the case $\lambda\neq 0$, where it is applied to the study of
supersymmetry algebras in $AdS^p \times S^q$ spacetimes). From Killing
spinors then we construct conformal Yano tensors, and show they are
spanned by  products of Killing vectors. This is the maximum
possible freedom allowed by the underlying structure.

\subsection{Geometry of symmetric spaces \label{sec:maxim_symm}} 
Let $M$ be a (connected) symmetric spacetime of
dimension $D$. Let a Lie group $G$ be its isometry group, then $dim(G)=D+h$,
$h\geq 0$. There is a natural action of $G$ on
$M$ and the isotropy group $H$, $dim(H)=h$, is a Lie
group as well. $M$ can  be
described as the homogeneous space $G/H$. The Lie algebra of $G$
can be written as $\mathbf{g}= \left\{ T_I\right\}=\left\{ M_i ,
P_a\right\}$, $I=1,\dots ,D+h$, $i=1, \dots h , \, a=1, \dots ,D$,
where the $M_i$ are generators of $\mathbf{h}$, the Lie algebra of
$H$, and the $P_a$ span a complementary algebra
$\mathbf{m}$. $\mathbf{h}$ and $\mathbf{m}$ are eigenspaces of a Lie
algebra involution with eigenvalues $+1$ and $-1$ respectively. Then
one can show that they have following commutation rules 
\begin{equation} 
	\begin{array}{lcl} 
	\left[ M_i , M_j \right] &=& f_{ij}^k M_k , \\
	\left[ M_i , P_a \right] &=& f_{ia}^b P_b , \\ 
	\left[ P_a , P_b \right] &=& f_{ab}^i M_i . 
	\end{array} 
	\label{eq:commutators} 
\end{equation} 
The first condition says that $\mathbf{h}$ is a subalgebra of
$\mathbf{g}$. The second one that $\mathbf{m}$ is a representation of
$\mathbf{h}$. The third one that commutators in $\mathbf{m}$ cannot
close in $\mathbf{m}$ itself. The spaces $\mathbf{h}$ and $\mathbf{m}$
are orthogonal with respect to the Killing metric. 
 
A generic element in $G/H$ can be obtained by exponentiating the
generators $P_a$: 
\begin{equation} 
	j(x)= e^{x^1 P_1}\dots e^{x^D P_D} , 
	\label{eq:representative}
\end{equation} 
and if a matrix representation for the algebra $\mathbf{g}$ is used, 
\begin{equation} 
	\begin{array}{lcl} 
	 M_i  &\rightarrow& R(M_i) , \\
	P_a &\rightarrow& R(P_a) , \\ 
	\end{array} 
\end{equation} 
then a representative $R(j(x))$ is obtained by exponentiating the
representatives of the $P_a$. Under a constant $g\in G$ transformation 
the element $g\, j(x)$ is no longer a coset representative of the form
\eref{eq:representative}, unless one applies a compensating
multiplication by $h\in H$ on the right, where $h$ is a function of
both $g$ and the coordinates $x$: 
\begin{equation} 
	j(x^\prime )= g\, j(x) h^{-1} . 
	\label{eq:coset_transf}  
\end{equation} 
 
\noindent An important part of the construction is the adjoint action of
$\mathbf{g}$ on any representation 
\begin{equation} 
	Adj_{T_I} (R(T_J))= R(T_K)\, R_{Adj}(T_I)^K_{\;\; J} :=
	\left[ R(T_I) , R(T_J) \right] , 
\end{equation} 
which when exponentiated 
\begin{equation} 
	 R_{Adj}(g(y))^K_{\;\; J} = e^{y^I \, R_{Adj}(T_I)^K_{\;\; J}}  
\end{equation} 
(where $y^I$ are coordinates for $G$), gives the adjoint action of $G$
on its Lie algebra 
\begin{equation} 
	R(g)\, R(T_I) \, R(g^{-1}) = R(T_J)\, R_{Adj}(g)^J_{\;\; I} . 
\end{equation} 
If it possible to define duals $T^I$ of the generators by some action
$T^I(T_J) = \delta^I_J$, for example when the
bi-invariant Killing metric 
\begin{equation} 
	K_{IJ} = Tr \left[ R(T_I) \, R(T_J) \right]  
\end{equation} 
is non degenerate, then these will transform under the adjoint action
of $G$ as 
\begin{equation} 
	R(g)\, R(T^I) \, R(g^{-1}) =  R_{Adj}(g^{-1})^I_{\;\; J}\, T^J
	. 
	\label{eq:adjoint_action} 
\end{equation} 
 
The last notion we need to introduce is that of a global frame for $M$
and its associated spin connection. The Maurer-Cartan $1$--form
$\Theta =  g^{-1}\, dg$ is left invariant and takes values in $\mathbf{g}$. It
provides a global frame for $G$. One can decompose it according to the
$G/H$ structure as: 
\begin{equation} 
	g^{-1}\, dg = e^a \, P_a + \theta^i \, M_i . 
\end{equation} 
$\left\{ e^a \right\}$ provides a global frame for $M$ and projecting
the Maurer-Cartan  equation $d\Theta + \Theta
\wedge \Theta$ on the $a$ components one finds 
\begin{equation} 
	d e^a \,  + \theta^i \wedge e^b \, f_{ib}^{a} = 0,  
\end{equation} 
which is the equation for a torsionless connection given by 
\begin{equation} 
	w^a_{\;\; b} = \theta^i \, f_{ib}^a . 
	\label{eq:connection} 
\end{equation} 
This is in agreement with the fact that under a transformation
eq.\eref{eq:coset_transf} $\theta$ transforms like a connection gauging
local $H$ transformations: 
\begin{equation} 
	\theta^i (x^\prime )= (h\,\theta(x)\, h^{-1})^i + (h^{-1}\,
	dh)^i . 
\end{equation} 
Given a constant bilinear form $N_{(ab)}$ which is non degenerate one
can construct  the metric 
\begin{equation} 
	ds^2= N_{ab} \, e^a \otimes e^b . 
\end{equation}
If it satisfies 
\begin{equation} 
	f_{i (a}^c \, N_{b)c} = 0 
\end{equation}
then it is left invariant under the action of the full group $G$ (this
is true for the projection of the Killing
metric $K_{IJ}$ on $a,b$ indexes when it is non degenerate) and it
admits at least $G$ as isometry group. Its associated Killing vectors
can be seen to be given by 
\begin{equation} 
	k_{(I)}^{\;\; a}(x) = R_{Adj}(j^{-1}(x))^a_{\;\; I} . 
	\label{eq:Killing_vectors}  
\end{equation}

\subsection{Yano tensors on maximally symmetric spaces
	\label{sec:special_tensors}} 
In this section we use the procedure of \cite{Ortin} to construct
Killing spinors on maximally symmetric spaces. From these we derive
Yano tensors. 
 
\noindent Start from the Killing equation 
\begin{equation} 
	0= ( \nabla -i\, \lambda \,\Gamma )\eta^+ = ( d
	+\frac{1}{4}\omega_{ab}\Gamma^{ab} -i\,
	\lambda \,\Gamma )\eta^+ . 
\end{equation} 
An explicit check shows that 
\begin{equation} 
	R_s(M_i) := \frac{1}{4}f_{ia}^b\, \Gamma_b^{\;\; a} 
\end{equation} 
provides a spinorial representation of $\mathbf{h}$ so that, using
\eref{eq:connection}, one can rewrite the Killing equation as  
\begin{equation} 
	0= ( d +\theta^i \, R_s(M_i)  -i\,\lambda \, e^a\, \Gamma_a )\eta^+ . 
\end{equation} 
Now suppose that the matrix $-i\,\lambda \Gamma_a$ provides a
representation of the generators $P_a$: 
\begin{equation} 
	R_s(P_a) := -i\,\lambda \Gamma_a , 
	\label{eq:assumption} 
\end{equation}
then the Killing equation can be put in the suggestive form 
\begin{equation} 
	\begin{array}{lcl} 
	0&=& ( d +\theta^i \, R_s(M_i) + e^a\, R_s(P_a) )\eta^+ = (d +
	R_s(j^{-1})\, d\, R_s(j) )\eta^+ \\ 
	&=& R_s(j^{-1}) d ( R_s(j)\, \eta^+ ) . 
	\end{array} 
	\label{eq:suggestive} 
\end{equation} 
This is solved by 
\begin{eqnarray} 
	\eta^{+ \alpha}&=& R_s(j^{-1})^\alpha_{\;\;\beta}\,
	\epsilon^\beta , \\ 
	R_s(j(x))&=& e^{x^1 R_s(P_1)}\dots e^{x^D R_s(P_D)} , 
\end{eqnarray} 
where $\epsilon$ is a constant spinor. 

\noindent As can be seen, the number of Killing spinors admitted is
maximal and a basis for them is given by 
\begin{equation} 
	\eta_{(\alpha )}^{+ \beta}= R_s(j^{-1})^\beta_{\;\;\alpha} . 
	\label{eq:sol_plus}   
\end{equation} 
An immediate consequence is that, since the Gamma matrices are
invariant tensors of $Spin(D)$ (or $Spin(1,D-1)$, depending on the
signature), then the basis transforms covariantly under $Spin(D)$
($Spin(1,D-1)$) rotations: 
\begin{equation} 
	\eta_{(\gamma )}^{+ \beta}(x^\mu ) U^\gamma_{\;\;\alpha} = 
	U^{\beta}_{\;\;\gamma} \eta_{(\alpha )}^{+
	\gamma}(\Lambda^{(-1)\mu}_{\hspace{.8cm}\nu} x^\nu ) ,  
	\label{eq:spinor_action}  
\end{equation} 
where $U$ is the  operator associated to the $SO(D)$
($SO(1,D-1)$) rotation $\Lambda^\mu_{\;\;\nu}$. 
 
\noindent Spacetimes like for example Sasaki-Einstein, almost K\"{a}hler,
almost $G_2$ mentioned in the previous section admit less than the
maximum number of Killing spinors and therefore are not described by
this approach. The reason is that one of the assumptions we made is
that $-i \lambda \Gamma_a$ is a representative of the $P_a$ generators
but, as we will see later, this does not happen in general.

Now we are in the position to  discuss the structure of special
tensors on these spaces. We restrict our attention to the Euclidean
signature case and $\lambda$ real, which encodes all the interesting
structure. The only differences one encounters
when changing to Lorentzian signature and\textbackslash or $\lambda$
imaginary are 
those related to the fact that the hermitian conjugate of the matrix
$R(P_a)$ will flip sign if $\lambda$ becomes imaginary and will be
sandwiched between two $\Gamma^0$ matrices in Lorentzian
signature. This is paired to the fact the tensors of kind $A$ and $B$
of the previous section interchange their roles. 

\noindent First of all let us make contact with
\eref{eq:normalization} that says it is possible to normalize the
spinors. $(+)$--type solutions of the Killing equation are given by
\eref{eq:sol_plus}. Solutions of $(-)$--type are obtained by changing
sign to $\lambda$ or, equivalently, by sending $R_s(P_a) \rightarrow -
R_s (P_a) := R_{-s} (P_a)$. This defines an equivalent representation
of the $P_a$ generators which is on the same footing as the first
one. If $\lambda$ is real then $R_s
(j^{-1})^\dag = R_s (j)$, from the definition
\eref{eq:assumption}. Therefore 
\begin{equation} 
	\overline{\eta}^{\alpha +}\eta^+_\beta = \left[ R_s (j) R_s
	(j^{-1}) \right]^\alpha_\beta = \delta^\alpha_\beta . 
\end{equation} 
If instead $\lambda$ is imaginary then this is no longer true and it
holds instead $R_{-s} (j^{-1})^\dag = R_s (j)$, from which the need to
put together $(+)$ and $(-)$ spinors in order to form constants. The
same holds in Lorentzian signature with minor changes. 
 
\noindent Now consider the special tensors $A$ and $B$. Write 
\begin{equation} 
	\begin{array}{lcl} 
	B^{(n)(\alpha )}_{(\beta )} &:=&      
	B^{(n)(\alpha )}_{(\beta ) [\mu_1\dots\mu_n ]} dx^{\mu_1}\dots
	dx^{\mu_n}   =
	\overline{\eta}^{\alpha +}  \, \Gamma^{a_1\dots a_n} \eta^+_{\beta} \,
	e_{a_1}\dots e_{a_n}  \\ &=& \left( -i \lambda\, D\right)^n
	R_s (j) R_s(P^{[a_1})\dots R_s(P^{a_n]}) R_s(j^{-1})
	e_{a_1}\dots e_{a_n} , 
	\end{array} 
\end{equation} 
where we defined representatives of the dual generators $R_s(P^a):=
\frac{i}{\lambda\, D} \Gamma^a$ according to 
\begin{equation} 
	Tr\left[ R_s(P_a)\, R_s(P^b) \right] = \delta_a^b . 
\end{equation} 
We can immediately recognize the adjoint action of $G$ on the dual of
$\mathbf{m}$, given by eq.\eref{eq:adjoint_action}. This allows to
rewrite $B$ as 
\begin{equation} 
	\left( -i \lambda\, D\right)^n R_{Adj}(j^{-1})^{[a_1}_{J_1}\,
	R_s(T^{J_1})^\alpha_{\;\;\lambda_1} \dots
	R_{Adj}(j^{-1})^{a_n]}_{J_n}\,
	R_s(T^{J_n})^{\lambda_{2n-2}}_{\;\;\beta} e_{a_1}\dots e_{a_n}
	.  
\end{equation}   
Now we just need to recall that Killing forms can be expressed as in
eq.\eref{eq:Killing_vectors}, and we get the final form 
\begin{equation} 
	B^{(n)(\alpha )}_{(\beta )} = \left( -i \lambda\, D \right)^n \left[
	M^{[J_1\dots J_n]} \right]^\alpha_{\;\;\beta} \,
	k_{(J_1)}\dots k_{(J_n)} , 
\end{equation} 
where $\left[ M^{[J_1\dots J_n]} \right]^\alpha_{\;\;\beta}$ is the constant
matrix 
\begin{equation} 
	\left[ M^{[J_1\dots J_n]} \right]^\alpha_{\;\;\beta} :=
	R_s(T^{[J_1})^\alpha_{\;\;\lambda_1} \dots
	R_s(T^{J_n]})^{\lambda_{2n-2}}_{\;\;\beta}  . 
\end{equation} 
Therefore we learn that $B$ tensors on these spaces have an extremely
simple form which is dictated by the large amount of symmetry: they
are spanned by products of Killing forms with constant coefficients. 
These tensors as well as their
spinorial building blocks transform in a
representation of $Spin(D)$, as it can be easily seen by
\eref{eq:spinor_action}. 
 
\noindent For what concerns tensors of kind $A$ almost nothing
changes: one has just to notice that $\overline{\eta}^{\alpha -}=
R_{-s} (j^{-1})^\dag \neq R_s (j)$, so that one gets the same
expression as above but multiplied by a prefactor 
\begin{equation} 
	\begin{array}{lcl}
	A^{(n)(\alpha )}_{(\beta )} &=& \left( -i \lambda\, D \right)^n \left[
	R_{-s}(j^{-1})^\dag \, R_s(j^{-1}) \right]^\alpha_{\;\;\gamma}
	\, \left[M^{[J_1\dots J_n]} \right]^\gamma_{\;\;\beta} \,
	k_{(J_1)}\dots k_{(J_n)} \\ &=& \left[
	R_{-s}(j^{-1})^\dag \, R_s(j^{-1}) \right]^\alpha_{\;\;\gamma}
	B^{(n)(\gamma )}_{(\beta )} , 
	\end{array}  
\end{equation}
that is, $A$ and $B$ tensors are related by an invertible, although
position dependent, transformation, and are equivalent. 
 
\noindent All the discussion easily carries over the Lorentzian case,
where simply the roles of $A$ and $B$ tensors are exchanged.

We conclude this session with a discussion of the assumption
\eref{eq:assumption}. If
the Gamma matrices define a representation of the generators $P_a$
then they must satisfy 
\begin{equation} 
	\left[ R_s(P_a), \, R_s(P_b) \right] = f_{ab}^i \, R_s(M_i) = 
	\frac{1}{4}\, f_{ab}^i \, f_{ic}^d \, \Gamma_d^{\;\; c} . 
\end{equation} 
At the same time, the definition \eref{eq:assumption} implies 
\begin{equation} 
	\left[ R_s(P_a), \, R_s(P_b) \right] = - 2\, \lambda^2 \,
	\Gamma_{ab} ,  
\end{equation}  
and these two conditions are consistent if 
\begin{equation} 
	f_{ab}^i \, f_{ic}^d \, = -8\,\lambda^2 \, \delta^d_{[a}\,
	 N_{b]c} . 
	\label{eq:cond1}  
\end{equation} 
In case the Killing form is invertible it is possible to massage
further this equation: multiply on the right by
$f_{jd}^c$  and use 
\begin{equation} 
	K_{ij} = f_{ic}^d \, f_{jd}^c + f_{il}^m \, f_{jm}^l = 
	f_{ic}^d \, f_{jd}^c + K^H_{ij} , 
	\label{eq:remainder} 
\end{equation} 
where $K^H$ is the Killing form of the group $H$. A little algebra
shows that \eref{eq:cond1} amounts to 
\begin{equation} 
	\left\{ \begin{array}{lcl} 
		K^{H}_{ij} &=& \alpha K_{ij} , \\ 
		K_{ab} &=& \frac{8\lambda^2}{1-\alpha} \, N_{ab} ,  
		\end{array}  \right.  
	\label{eq:cond2} 
\end{equation} 
where $\alpha$ is a non determined proportionality constant. 
These conditions turn out to leave some freedom in the choice of $G$
and $H$, and therefore in the topology of the spacetime, but are quite
restrictive from the point of view of the metrics admitted, as we are
going to show below. 

\noindent The first condition is group-theoretical. It
says that the projection on $\mathbf{h}$ of the Killing form of $G$
must be proportional to that of $H$ itself. This restricts the set of
possible groups $G$ and $H$. The geometrical meaning is the
following. Using the Jacobi identity with respect to the triple of
indexes $(i,j,a)$ and the commutation relations
eq.\eref{eq:commutators} one finds that the matrices $\left( M_i
\right)^a_{\;\; b} := f_{ib}^a $ are a representation of the algebra
$\mathbf{h}$:  
\begin{equation} 
	\left[ M_i , M_j \right] = f_{ij}^k \, M_k . 
	\label{eq:representation} 
\end{equation} 
From eq.\eref{eq:remainder} it is seen that the difference between the
Killing form (restricted on $\mathbf{h}$) $K_{ij}$ and the Killing
form of $H$ $K^H_{ij}$ is given exactly by the trace of the
representation of eq.\eref{eq:representation}. What we are asking is
that this trace, which by definition is the Killing form of the
representation, is proportional to $K^H_{ij}$ itself which is the
Killing form of the adjoint representation. This is
guaranteed if $H$ is simple, in which case  its Lie algebra is
irreducible, and therefore the Killing forms of all representations
come out to be proportional to each other. If $H$ is semi-simple
instead the argument  may fail
since the adjoint representation is faithful but reducible, and
therefore in order for the condition to be satisfied one has to
carefully pick up the coefficients $f_{ib}^a$ so that they define a
representation which is not irreducible but made up of the same direct
sum of representations as the adjoint one. 
 
\noindent To summarize, the condition is certainly satifisfied by
simple groups $H$ like in the case of the $D$-dimensional sphere $S^D
= SO(D+1)/SO(D)$,  the hyperbolic plane $H^D = SO(1,D)/SO(D)$ and their 
Lorentzian analogues de Sitter space $dS^D = SO(1,D)/SO(1,D-1)$ and
anti-de Sitter $AdS^D = SO(2,D-1)/SO(1,D-1)$. The same argument holds
if one orbifolds such manifolds by taking the quotient with a finite
group $\mathbb{Z}_n$, since condition \eref{eq:cond2} only
affects the continuous (Lie group) part of the algebra. Therefore
real projective spaces $\mathbb{P}^D(\mathbb{R} ) = S^D/\mathbb{Z}_2$
(which are spin manifolds for $D$ odd) and all further quotients
$S^D/\mathbb{Z}_n$, $H^D/\mathbb{Z}_n$ and their Lorentzian analogues
apply to our discussion. A counterexample where $H$ is not simple but
just semi-simple and 
\eref{eq:cond1} is still satisfied is given by complex projective
spaces  $\mathbb{CP}^n = U(n+1)/U(n)\times U(1)$
(excluding the case $n$ even for which the manifolds do not admit a spin
structure). 

The second condition concerns the metric of the
space. This has to be proportional to the projection of the Killing
form on $\mathbf{m}$ indices and, once the constant $\alpha$ is known,
different constants of proportionality will lead to different values
of the constant $\lambda$ appearing in the Killing equation. The
standard metric on the spaces discussed above satisfies the condition,
but for example the Fubini-Study metric on $\mathbb{CP}^n$, $n>1$,
does not. Hence, once the topology if fixed by the $G/H$ structure,
only the standard metric on the manifold can be used.

At this point the reader may be confused by the following fact. Even
if we restrict our attention to the simplest case of $\lambda$ real
and Riemannian manifolds, it seems that, as long as we find an
appropriate quotient $G/H$ satisfying condition \eref{eq:cond2}, then
we can put the standard metric on it and get a manifold with the
maximum number of Killing spinors! This seems to disagree with the
fact one would expect only the $D$-dimensional sphere to admit maximum
number of Killing spinors. The solution to this apparent puzzle, as
mentioned in \ref{sec:almost_special_hol}, is that maximum number of
Killing spinors is associated to a flat cone over the manifold, and for
this to happen only maximal symmetry is necessary.
For example,
real and complex projective spaces are maximally symmetric and the
former is not even simply connected. Alternatively, using an orbifold
$S^D/\mathbb{Z}_n$ one discover that its cone is a proper cone (in the
literal sense), i.e. a manifold with $\mathbb{Z}_n$ holonomy which is
everywhere flat but on its tip, where the Riemann tensor picks up a
$\delta$--like singularity. Not by chance the tip of the cone is set
at $r=0$, where $r$ is the radial coordinate of the cone. One can also
take the cone over a complex projective space $\mathbb{CP}_n$ and
still get a space which is everywhere flat but on the origin. The only
requirement one has to satisfy in order to get a cone which is flat
out of the origin is that the base manifold has, locally, the metric
of a maximally symmetric space.

\noindent As a final remark we comment on another, different case
which is not contemplated in this
approach: homogeneous pp--wave
spacetimes. They are symmetric spacetimes but the Lie algebra of their
isometry group contains a central element. Therefore the Killing
form is not invertible and another metric $N_{ab}$ has to be
used. This implies the representatives of $P_a$ generators cannot be
taken to be simply Gamma matrices. As a matter of fact Killing spinors
on these spacetimes  are not present in the maximum number and therefore cannot
be obtained using the procedure outlined in this section. Rather, it
can happen instead that, when the pp--wave spacetime is a maximally
supersymmetric solution of supergravity, then  it is a generalized
kind of Killing spinors that
appears in maximal number. These correspond to spinors associated to
 transformations of supergravity fields in presence of fluxes, and
accordingly to this the representatives of $P_a$ generators are not
given by eq.\eref{eq:assumption}, but a different definition where the
flux appears (see \cite{Ortin} for an example with $\mathcal{N} =1$ in
$D=11$ dimensions). The presence of fluxes can give rise to new kind
of special tensors on these manifolds, like calibrations, and we
reserve their study to
future communications. Yano tensors of rank $2$ and $3$ on a
generalized pp--wave background have been studied in
\cite{BaleanuBaskalThanksGuys!!!}.

\section{Conclusions \label{section:conclusions}} 
The main purpose of this paper was to study in detail the quantum mechanical role
of Yano tensors. We started from the theory of the (quantized)
spinning particle and from there have shown that they allow to define operators
which commute with the Dirac operator in curved spacetime, for every
possible rank of the tensor used. These operators provide quantum numbers
for the Dirac equation even when isometries are completely absent. One
possible application of this is the study of the Dirac equation on
nearly Calabi-Yau or $G_2$ manifolds, since they have almost special
holonomy. Being linear in the momenta, the operators define genuine
dynamical symmetries of the system, i.e. symmetries of the phase
space, and could be relevant in the definition of representations of the
quantum group of dynamical symmetries on a given spacetime. 
 
Manifolds of special holonomy in general admit Yano and conformal Yano
tensors. We have constructed them explicitly for maximally symmetric
spaces, showing that their form is extremely simple.

We believe that Yano operators deserve further attention. They are
related to a mulititude of different topics like exotic
supersymmetries, classes of manifolds which are more general
than K\"{a}hler ones, index theorems, supergravity theories.

\acknowledgments{I am grateful to Gary Gibbons for drawing my
attention to Yano tensors and for illuminating discussions. Many
thanks to Sean Hartnoll, Sigbjorn Hervik, Jasbir Nagi, Carlos
Nu$\tilde{\mbox{n}}$ez, Ruben Portugues, Guillermo Silva for useful
suggestions  along different
stages of this work, and for cups of coffee on the roof of CMS. The
author is funded by EPSRC, Fondazione
``A. Della Riccia'', Firenze, and Fondazione ``Ing. A. Gini'', Padova.}

\section*{Appendix 1: Proof of the Lemma}
\addcontentsline{toc}{section}{\bf Appendix}  
The (anti)commutator splits into a classical and a quantum part. The
classical part vanishes due to the result of
\cite{GibbonsRietdijkVanHolten} and \cite{Tanimoto}. So we just
need to check that quantum corrections vanish. 

\noindent There are two sources of ``anomaly'' in the quantum
corrections. One is related to the fact that the Clifford algebra of Gamma
matrices is different from the Grassmann algebra of spin coordinates
$\psi^\mu$, eq.\eref{eq:anticomm}. The second is the well known
operator ordering problem  due to the
quantization procedure $\Pi_\mu \rightarrow \hbar D_\mu$. 
Direct inspection shows that contributions due to operator
ordering are of the same kind of those coming from Gamma matrix
(anti)commutators (their only source is given by commutators of
covariant derivatives).  One
can reorganize the expression $\left[  D\hspace{-0.22cm} /
\right. \left. , Y_r \right\}$ in terms of powers of Gamma
matrices. Order by order they are given by $\Gamma^{r+2}$,
$\Gamma^{r}$, $\Gamma^{r-2}$, $\dots$.  The term proportional to
$\Gamma^{r+2}$ has no contractions since is the one of highest
order, therefore is a classical term. Possible terms coming from
quantum corrections are those of
order $r$, $r-2$, ... One can easily see that the only terms allowed
were contractions appear are the following.

\noindent For order $r$:  
	\[ 
			\Gamma^{a_1\dots a_r} f_{[a_1\dots
	a_{r-2}}^{\hspace{1.2cm}b_1b_2} R_{a_{r-1}a_r]b_1 b_2} ,  
	\] 
	\[ 
			\Gamma^{a_1\dots a_r} f_{[a_1\dots
	a_{r-1}}^{\hspace{1.2cm}b} \left( Ric \right)_{a_r]b} .  
	\] 
For order $r-2$: 
	\[ 
			\Gamma^{a_1\dots a_{r-2}} f_{[a_1\dots
	a_{r-3}}^{\hspace{1.2cm}b_1b_2b_3} R_{a_{r-2}]b_1 b_2 b_3} = 0,  
	\] 
	\[
			\Gamma^{a_1\dots a_{r-2}} f_{a_1\dots
	a_{r-2}}^{\hspace{1.2cm}b_1b_2} \left( Ric \right)_{b_1 b_2} = 0 . 
	\]  
All orders from $r-2$ downwards vanish because there are no further nonzero
contractions. The term $f_{a_1\dots a_r}\, R$ cannot appear since is
of order $r$ because $R$ comes from the Riemann tensor after two
contractions\footnotemark . \footnotetext{Here we are talking about
two contractions appearing {\it algebraically} from Gamma matrices
commutators. Of course the term $f_{a_1\dots a_r}\, R$ can appear from
a term of order $r$  when
one contraction is provided from the Riemann tensor rather than an
anticommutator of Gamma matrices. For example, on
Einstein spaces, it will come from the term $f_{a_1\dots
	a_{r-1}}^{\hspace{1.2cm}b} \left( Ric \right)_{a_r b}$.
Therefore we consider this as special case of the discussion of order
$r$ terms.} 
 
\noindent Therefore all that one has to do is to check the order
$r$. A detailed calculation shows that
there appear two non trivial possible quantum corrections in the
(anti)commutator.  
The first is 
\begin{equation} 
	\frac{(-1)^{r-1}}{4} \Gamma^{\mu_1\dots \mu_r} f_{[\mu_1\dots
	\mu_{r-1}}^{\hspace{1.2cm}b}\, R_{\mu_r]b a_1 a_2} \Gamma^{a_1
	a_2} . 
\end{equation} 
Using the expansion 
\begin{equation} 
	\Gamma_{\mu_1\dots \mu_r}  \Gamma^{a_1a_2} =
	\Gamma^{a_1a_2}_{\hspace{.8cm}\mu_1\dots\mu_r} + 2\, \hbar \,r
	\, 
	\delta^{[a_1}_{[\mu_1 } \, \Gamma^{a_2]}_{\;\;\;
	\mu_2\dots\mu_r ]} + \left(\mbox{lower orders}\right) ,  
\end{equation} 
one can extract the order $r$ term and show it amounts to 
\begin{equation} 
	(-1)^{r+1}\left( -\frac{r-1}{4}\, \Gamma^{a_1\dots a_r} f_{[a_1\dots
	a_{r-2}}^{\hspace{1.2cm}b_1b_2} R_{a_{r-1}a_r]b_1 b_2} +
	\frac{1}{2} \, \Gamma^{a_1\dots a_r} f_{[a_1\dots
	a_{r-1}}^{\hspace{1.2cm}b} \left( Ric \right)_{a_r]b} \right)
	. \label{eq:first_contribution} 
\end{equation} 
The second non trivial part is given by 
\begin{equation} 
	\frac{(-1)^{r+1}}{2(r+1)}\, \Gamma^{\lambda} \, \Gamma^{\mu_1\dots
	\mu_{r+1}}  \nabla_{\lambda} \nabla_{\mu_1}  f_{\mu_2\dots
	\mu_{r+1}} . 
\end{equation} 
Using the integrability condition \eref{eq:int_cond} and the
identity 
\begin{equation} 
	\Gamma^{\lambda} \,  \Gamma_{\mu_1\dots \mu_{r+1}}  =
	\Gamma^{\lambda}_{\hspace{0.5cm}\mu_1\dots\mu_{r+1}} + \hbar
	\, (r+1)
	\, \delta^{\lambda}_{[\mu_1 } \, \Gamma_{
	\mu_2\dots\mu_{r+1} ]}  ,  
\end{equation} 
one can calculate the order $r$ contribution and see that it exactly
cancels the other term of eq.\eref{eq:first_contribution}. $\triangle$

\section*{Appendix 2: Hodge duality and conformal Yano equation} 
First of all decompose the covariant derivative of a rank $r$ form $f$
according to 
\begin{equation}
	\begin{array}{lcl} 
	\nabla_{ \mu_1} f_{\mu_2 \mu_3\dots \mu_{r+1} } &=&
	 \nabla_{[
	\mu_1} f_{\mu_2\mu_3  \dots \mu_{r+1}] } + \\ 
	&& \hspace{-1cm}\frac{2}{r+1} \left( \nabla_{(
	\mu_1} f_{\mu_2 )\mu_3 \dots \mu_{r+1} } + \nabla_{( \mu_1}
	f_{|\mu_2 | \mu_3 )\dots \mu_{r+1} }+ \dots + \nabla_{( \mu_1}
	f_{| \mu_2 \mu_3 \dots \mu_r | \mu_{r+1}) } \right) . 
	\end{array} 
	\label{eq:der_decomp} 
\end{equation} 
Let $f^*$ be the dual of $f$. Then the antisymmetric part of the
covariant derivative can be written as 
\begin{equation} 
	\nabla_{[\mu_1} f_{\mu_2 \mu_3 \dots \mu_{r+1}] } = \frac{
	(-1)^{D(r+1)+t} }{r+1} \left( * \delta f^* \right)_{\mu_1 \dots
	\mu_{r+1}} , 
	\label{eq:antisymm_der}   
\end{equation} 
using $**= (-1)^{r(D-r)+t}$, $\delta = (-1)^{D(r+1)+1+t} *d*$, where
$t$ is the number of time dimensions.   
Suppose now $t=0$ for simplicity. The discussion for $t=1$ is analogous. One can  
calculate $\nabla f^*$ using the decomposition
\eref{eq:der_decomp}, obtaining 
\begin{equation}
	\begin{array}{ll} 
	\nabla_{\mu_1} f^*_{\mu_2 \mu_3 \dots \mu_{(D-r+1)} } = 
	\frac{(-1)^{r(D-r)}}{r!} 
	\eta_{\mu_2 \mu_3 \dots \mu_{(D-r+1)}}^{\hspace{2.2cm} \lambda_1 
	\dots \lambda_r} \, \left[ \right. & \nabla_{[ \mu_1} f_{\lambda_1 \dots
	\lambda_{r}] } +      \\ & \left. 
  	\frac{2}{r+1} \left( \nabla_{( \mu_1} f_{\lambda_1 ) \dots
	\lambda_{r} } + 
	\dots \right) \right] . 
	\end{array}  \label{eq:calc_nabla_fstar} 
\end{equation} 
Now, the contribution to the totally antisymmetric part of $\nabla f^*$ comes only from 
the terms in round brackets. Considering only this part, and substituting 
eq.\eref{eq:conformal_Yano_def} in it, we get 
\begin{eqnarray} 
	&& 2 \, \frac{(-1)^{r(D-r)}}{(r+1)!} \, 
	\eta_{\mu_2 \mu_3 \dots \mu_{(D-r+1)}}^{\hspace{2.2cm} \lambda_1 
	\dots \lambda_r} \, \left[ g_{\mu_1 \lambda_1} \,
	\Phi_{\lambda_2 \dots \lambda_r} + 
	A \, g_{\lambda_2 ( \mu_1} \, \Phi_{\lambda_1 ) \lambda_3 \dots \lambda_r } \right] 
	\nonumber \\ &=& \frac{r \, (2-A)}{(r+1)!} (-1)^{r(D-r)} \, 
	\eta_{\mu_2 \mu_3 \dots \mu_{(D-r+1)} \mu_1}^{\hspace{1cm} \lambda_2 \dots \lambda_r} 
	\Phi_{\lambda_2 \lambda_3 \dots \lambda_r } \\ &=& (-1)^{r+1}
	\frac{2-A}{r+1} 
	*\Phi_{\mu_1 \dots \mu_{(D-r+1)}} =  
	(-1)^{r} \frac{2-A}{(D+A)(r+1)} \, *\delta f_{\mu_1 \dots
	\mu_{(D-r+1)}} . 
\end{eqnarray} 
On the other hand, substituting $f$ with $f^*$ in \eref{eq:antisymm_der} gives 
\begin{equation} 
	\begin{array}{lcl}
	\nabla_{[\mu_1} f^*_{\mu_2 \mu_3 \dots \mu_{D-r+1}] } &=& 
	\frac{(-1)^{D\left( (D-r)+1\right)}+t }{(D-r)+1}
	\left( * \delta *f^* \right)_{\mu_1 \dots\mu_{D-r+1}} \\ 
	&=& \frac{(-1)^{r} }{D-r+1} \left( * \delta f \right)_{\mu_1 \dots
	\mu_{D-r+1}}   , 
	\end{array} 
\end{equation} 
which is consistent with the previous expression if and only if $A= 1-r$. 
Once $A$ is fixed a
complete expression of $\nabla f^*$ can be calculated from
\eref{eq:calc_nabla_fstar} and the result is:
\begin{equation} 
	\nabla_{\mu_1} f^*_{\mu_2 \mu_3 \dots \mu_{D-r+1} } = 
	\nabla_{[\mu_1} f^*_{\mu_2 \mu_3 \dots \mu_{D-r+1}] } -
	\frac{D-r}{r+1} g_{\mu_1 [ \mu_2}\delta
	f^*_{\mu_3\dots\mu_{D-r+1}] } , 
\end{equation} 
which implies 
\begin{equation} 
	\hspace{-.01cm}\nabla_{(\mu_1} f^*_{\mu_2 ) \mu_3 \dots \mu_{D-r+1} }\hspace{-.1cm} = 
	\hspace{-.1cm} - \frac{1}{r+1}\hspace{-.1cm}\left( g_{\mu_1\mu_2}\delta
f^*_{\mu_3\dots\mu_{D-r+1} }\hspace{-.1cm} -\hspace{-.1cm}
	(D\hspace{-.1cm} -\hspace{-.1cm}  r \hspace{-.1cm}-\hspace{-.1cm}  1)g_{[\mu_3 |(\mu_1 }\delta
f^*_{\mu_2 )| \mu_4 \dots \mu_{D-r+1} } \right) ,  
\end{equation} 
that is in agreement with the conformal Yano equation
\eref{eq:conformal_Yano_def2} and the condition \linebreak $\Phi_{f^*} = -
\frac{1}{r+1}\delta f^*$. When the signature is Lorentzian the only difference is the 
appearence of some extra $(-1)$ terms. 
 
From what shown above we can easily see that the dual of a
Yano tensor $f$ is a closed conformal Yano tensor, and in particular
it is a Yano tensor itself if and only if both $f$ and $f^*$ are harmonic.

\section*{Appendix 3: Some examples of spaces of (almost) special holonomy} 
In this section we give some example of spaces with special or almost
special $SU(3)$ and $G_2$ holonomy, to which our construction
applies. They are either $G_2$ holonomy spaces which are
asymptotically locally conical (ALC) spaces over manifolds with almost
$SU(3)$ holonomy, or $Spin(7)$ spaces which are ALC  over manifolds with almost
$G_2$ holonomy.  
Such spaces are of direct relevance in compactifications of
String/$M$--Theory that preserve either $\mathcal{N}=1$ or
$\mathcal{N}=2$ supersymmetry, with or without fluxes. We refer in the
following to \cite{Boss} and \cite{BilalSfetsos}. 

\vspace{\baselineskip}\noindent
{\bf $G_2$ holonomy metrics} 

\vspace{\baselineskip}\noindent
Spaces of $G_2$ holonomy can arise in compactifying $M$--Theory down
to four dimensions while preserving $\mathcal{N}=1$ supersymmetry. It
is well known that compactification on a smooth $G_2$ manifold cannot
lead to chiral fermions (\cite{Witten}), so that an alternative
approach is to use singular $G_2$ manifolds. The metrics we present
here have $G_2$ holonomy and are locally asymptotic to conical
metrics. From the metric on the cone one can immediately read the metric
on the base manifold, which will be automatically of almost $SU(3)$
holonomy. In \cite{BilalMetzger} is shown how to associate, to every
such non-compact ALC manifold of $G_2$ holonomy, a compact manifold
with almost $G_2$ holonomy and conical singularities. This means that the
latter, close to the conical singularities, looks like the former ALC
manifold. Studying the Dirac equation  on such compact manifolds
using conserved quantities generated by Yano tensors may be
directly relevant in establishing the existence of chiral fermions,
as an approach that is complementary to those already used in the
literature \cite{Sean}. In all these cases the possible antisymmetric
tensors that can be defined by \eref{eq:A_tensor},\eref{eq:B_tensor}
and are non  zero are given by a $3$--form $\phi$ and its dual
$*\phi$. Since they 
are covariantly  constant, each of these is both a calibration for the
$G_2$ manifold and a Yano tensor. It would be interesting to see whether these 
forms, as well as in the rank $2$ case, can provide genuine String Theory 
symmetries. 
In the conical limit one recovers a metric of almost $SU(3)$ holonomy on the
base of the cone. Then the forms that are present on the base manifold
are a dimensional reduction of $\phi$, $*\phi$, and are an almost
complex structure $J$, a holomorphic volume form $\rho$ and
its dual $*\rho$. They satisfy the conditions $J_{\lambda\mu}\,
J^{\mu}_{\;\;\nu} = - g_{\lambda\nu}$, $d\, J \sim \rho$, $d*\rho \sim
J\, J$. $J$ is a Yano tensor of kind $A$, while $*\rho$ a Yano tensor
of kind $B$ and $\rho$ a conformal Yano tensor (see
\cite{Boss_BianchiIX} for a discussion of almost $SU(3)$ holonomy). 

Hence, the link between Yano tensors, special holonomy, supergravity
and calibrations is manifest.  
  
\vspace{\baselineskip}\noindent  
{\bf $G_2$ metrics over $SU(3)/U(1)\times U(1)$.} 

\vspace{\baselineskip}\noindent
Let $L_A^{\;\;B}$, $A,B=1,2,3$, be left invariant forms of $SU(3)$,
$L_A^{\;\;A} =0$, $(L_A^{\;\;B})^\dag = L_B^{\;\;A}$. Left
invariant forms for $SU(3)/U(1)\times U(1)$ are given by $\sigma =
L_1^{\;\; 3}$, $\Sigma= L_2^{\;\; 3}$, $\nu= L_1^{\;\; 2}$. They split
into real forms defined by $\sigma = \sigma_1 + i \sigma_2$, etc. 
Then the following metric has $G_2$ holonomy and principal orbits given by
$SU(3)/U(1)\times U(1)$: 
\begin{equation} 
	\begin{array}{lcl}  
	ds_7^2 &=& \frac{dr^2}{u\, v\, w}+ \frac{u\, w}{v}
	\left(\sigma_1^2 + \sigma_2^2\right) +  \frac{u\, v}{w}
	\left(\Sigma_1^2 + \Sigma_2^2\right) +  \frac{v\, w}{u}
	\left(\nu_1^2 + \nu_2^2\right) , \\ 
	u^2 &=& 4(r-r_1), \hspace{.75cm} v^2 = 4(r-r_2),  \hspace{.75cm} w^2
	= 4(r-r_3), 
	\end{array} 
\end{equation} 
where $r_1,r_2,r_3$ are constants of integration. The metric is
singular unless two of the $r_i$ are equal. 
The asymptotical
behaviour at $r\rightarrow +\infty$ is given by 
\begin{equation} 
	ds_7^2 = dR^2 + R^2 
	\left(\sigma_1^2 + \sigma_2^2 + \Sigma_1^2 + \Sigma_2^2 +
	\nu_1^2 + \nu_2^2\right) , 
\end{equation} 
where $R= \sqrt{2}\, r^{1/4}$. Then one obtains a metric of almost
$SU(3)$ holonomy on the base. The rank $3$ Yano tensor is given by 
\begin{equation} 
	\phi= \sqrt{u\, v\, w}\, Re \left(\overline{\sigma}\, \Sigma
	\, \nu \right) 
	+ i \left( - \frac{u\, w}{v} \sigma \, \overline{\sigma}  + 
	\frac{u\, v}{w} \Sigma \, \overline{\Sigma} + \frac{v\, w}{u} \nu \, 
	\overline{\nu}  \right) \, dt , 
\end{equation} 
where $dr=\sqrt{u\, v\, w} \, dt$.

\vspace{\baselineskip}\noindent  
{\bf $G_2$ metrics over $\mathbb{CP}^3$.} 

\vspace{\baselineskip}\noindent
Take $L_{AB}$, $L_{AB}=-L_{BA}$, $A=0,1,\dots 4$, to be left invariant forms of
$SO(5)$. Split the index $A$ into $A=(a,4)$, $a=0,\dots 3$, and define
$P_a  := L_{a4}$, $R_i :=
\frac{1}{2}(L_{0i}+\frac{1}{2}\epsilon_{ijk}\, L_{jk})$, $L_i :=
\frac{1}{2}(L_{0i}-\frac{1}{2}\epsilon_{ijk}\, L_{jk})$, $i=1,2,3$. 
Then one can write 
\begin{equation} 
	\mathbb{CP}^3= \frac{SO(5)}{SU(2)_L \times U(1)_R} , 
\end{equation} 
where $SU(2)_L$ is generated by the forms $L_i$, and $U(1)_R$ by
$R_3$. The following metric has $G_2$ holonomy and $\mathbb{CP}^3$
principal orbits: 
\begin{equation} 
	ds_7^2 = \left( 1- \frac{l^4}{r^4}\right)^{-1} \, dr^2
	+ r^2 \, \left( 1- \frac{l^4}{r^4}\right)\, \left(
	R_1^2 + R_2^2 \right) + \frac{r^2}{2} \, P_a^2 ,  
\end{equation} 
and $l$ is a constant. The conical limit is evident. 
The Yano tensor $\phi$ can be written as 
\begin{eqnarray} 
	\phi &=& dt \, \left[ a^2 R_1 \, R_2 - b^2 \left( P_0 \, P_3 + P_1 \, P_2 \right) 
	\right] \\ 
	&& + a \, b^2 \left[ R_2 \, \left( P_0 \, P_1 + P_2 \, P_3 \right) - 
	R_1 \, \left( P_0 \, P_2 + P_3 \, P_1 \right) \right] , 
\end{eqnarray} 
where $dr=\sqrt{1- l^4/r^4} \, dt$ and the functions $a(t)$, $b(t)$ obey the first order 
equations $\dot{a} = \frac{1}{2} \, a^2 \, b^{-2}$ and $\dot{b} = - \frac{1}{2} \, a \, 
b^{-1}$.

\vspace{\baselineskip}\noindent  
{\bf $G_2$ metrics over $S^3\times S^3$.} 

\vspace{\baselineskip}\noindent
This time the base manifold will be $S^3\times S^3= SU(2)\times
SU(2)/\mathbb{Z}_2 $. Let $\sigma_i$, $\Sigma_i$, $i=1,2,3$,
left-invariant forms associated to the two copies of $SU(2)$. A metric
with $G_2$ holonomy is given by 
\begin{equation} 
	ds_7^2 = dr^2 + a_i^2(r)\left( \sigma_i - \Sigma_i\right)^2 + 
	b_i^2(r)\left( \sigma_i +  \Sigma_i\right)^2 . 
\end{equation} 
$\mathbb{Z}_2$ symmetry is associated to the exchange of the two
copies of $SU(2)$. The functions $a_i$, $b_i$ have to satisfy a set of
differential equations explicitly given in \cite{Boss}, pg.46. For
what concerns the behaviour at infinity, a subset of all possible
solutions is given by those satisfying $a_2 = a_3$, $b_2 = b_3$. These
are all regular solutions and at $r\rightarrow +\infty$ behave as 
\begin{equation} 
	ds_7^2 \sim dr^2 + \frac{r^2}{4} \left(  \left( \sigma_1 -
	\Sigma_1\right)^2  + \frac{3}{4}\, \left( \sigma_2
	-\Sigma_2\right)^2 + \frac{3}{4} \,  \left( \sigma_2 +
	\Sigma_2\right)^2 \right) . 
\end{equation} 
The Yano tensor for the general metric is 
\begin{equation}
	\begin{array}{lcl} 
	\phi &=& a_1 a_2 a_3 \, (\sigma_1 - \Sigma_1)((\sigma_2 - \Sigma_2)(\sigma_3 - 
	\Sigma_3) - a_1 b_2 b_3 \, (\sigma_1 - \Sigma_1)((\sigma_2 + \Sigma_2)(\sigma_3 + 
	\Sigma_3) \\ &+& a_2 b_1 b_3 \, (\sigma_2 - \Sigma_2)((\sigma_1 + \Sigma_1)(\sigma_3 + 
	\Sigma_3) - a_3 b_1 b_2 \, (\sigma_3 - \Sigma_3)((\sigma_1 + \Sigma_1)(\sigma_2 + 
	\Sigma_2) \\ &+& dr \,  a_i b_i \, \sigma_i \Sigma_i . 
	\end{array} 
\end{equation}.

\vspace{\baselineskip}\noindent
{\bf $Spin(7)$ holonomy metrics} 

\vspace{\baselineskip}\noindent
Here we contemplate two cases of spaces with Spin$(7)$
holonomy. Almost Spin$(7)$ holonomy does not exist since there are no
possible cones of dimension $9$ that admit covariantly constant
spinors. In the asymptotically conical limit one recovers almost $G_2$
metrics on the base manifolds. The only non zero $p$--form is a
self-dual $4$--form $\Omega$, which is both a calibration and a Yano
tensor. On the base of the cone, by dimensional reduction and
consistently with almost $G_2$ holonomy, one recovers a $3$--form $\phi$
and a $4$--form $*\phi$. $\phi$ is a type $B$ Yano tensor that
satisfies $d\phi \sim *\phi$, and is a calibration, while $*\phi$ is
type $B$ conformal Yano tensor such that $d*\phi = 0$. The common
feature of all the cases discussed so far is that calibrations are
always Yano tensors, whether flux is or is not turned on.

\vspace{\baselineskip}\noindent  
{\bf $Spin(7)$ metrics over $SO(5)/SU(2)$.} 

\vspace{\baselineskip}\noindent
Here $SO(5)/SU(2)_L = S^7$ is defined in the same way as in the case of
$G_2$ metrics over $\mathbb{CP}^3$, but without taking the quotient
with respect to the $U(1)_R$ generator $R_3$. A metric with Spin$(7)$
holonomy is given by 
\begin{equation} 
	ds_8^2 = dr^2 + a_i(r)^2 \, R_i^2 + b(r)^2 \, P_a^2 , 
\end{equation} 
where the functions $a_i$, $b$ satisfy the differential equations 
\begin{equation} 
	\begin{array}{lcl} 
		\dot{a}_1 &=& \frac{a_1^2 - (a_2 - a_3)^2}{a_2 \, a_3} -
		\frac{a_1^2}{2\, b^2} , \hspace{.5cm} \mbox{and cyclic} , \\ 
		\dot{b} &=& \frac{1}{4} \, \sum_{i=1}^3 \frac{a_i}{b} . 
	\end{array} 
\end{equation} 
One particular solution is given by 
\begin{equation} 
	\begin{array}{lcl} 
		a_i = - \frac{\lambda}{4} \, r \, A_i , && b= -
		\frac{\lambda}{4} \, r \, B , 
	\end{array} 
\end{equation} 
with 
\begin{equation} 
	\begin{array}{lcl} 
		\lambda \, A_1 &=& -4 \, \frac{A_1^2 - (A_2 -
		A_3)^2}{A_2 \, A_3}  + 2\, \frac{A_1^2}{B^2} ,
		\hspace{.5cm} \mbox{and cyclic}  , \\ 
		\lambda \, B^2 &=& -  \sum_{i=1}^3 A_i ,  
	\end{array} 
\end{equation} 
and it is particularly easy to see the conical limit in this case. Since it is of 
cohomogeneity one, it is easy to display a rank four Yano tensor: it is constructed from the 
vielbein $e^A$ as 
\begin{equation} 
	\Omega = \frac{1}{4!} \, \Psi_{ABCD} \, e^A e^B e^C e^D , 
\end{equation} 
where $\Psi_{ABCD}$ is a $Spin(7)$ tensor constructed as following: 
\begin{eqnarray} 
	\Psi_{a b c 8} &=& \psi_{abc} , \\ 
	\Psi_{abcd} &=& \epsilon_{m_1 m_2 m_3 abcd} \, \psi_{m_1 m_2 m_3} , 
\end{eqnarray} 
where the index $a,b$ ranges from $1$ to $7$ and $\psi_{abc}$ is the $G_2$ invariant 
antisymmetric tensor whose non zero components are given by 
\begin{equation} 
	\psi_{123} = \psi_{516} = \psi_{624} = \psi_{435} = \psi_{471} = \psi_{673} = 
	\psi_{572} = 1 . 
\end{equation}  
$\Omega$ decomposes over the almost $G_2$ base as 
\begin{equation} 
\Omega = (\frac{\lambda r}{4})^3 \, dr \, \phi + (\frac{\lambda r}{4})^4 \, * \phi , 
\end{equation} 
such that 
\begin{equation} 
	\tilde{d} \phi = \lambda * \phi . 
\end{equation}, 
where $\tilde{d}$ is the differential operator on the base. 
 
\vspace{\baselineskip}\noindent  
{\bf $Spin(7)$ metrics over $SU(3)/U(1)$.} 

\vspace{\baselineskip}\noindent
In this case principal orbits are given by Aloff-Wallach spaces. These
are spaces of the form $SU(3)/U(1)$, where
$U(1)=diag(e^{ik},e^{il},e^{-i(k+l)})$. When $k,l$ are relatively
prime the
spaces are simply connected and denoted by $N(k,l)$. Almost $G_2$
metrics on these spaces are less interesting since they are smooth,
without singularities. Take $SU(3)/U(1)\times U(1)$ generators as in
the above example of  $G_2$ metrics over the space,
and add the generator $\lambda = \sqrt{2} \, \mbox{cos}\tilde{\delta} \,
L_1^{\;\; 1} +  \sqrt{2} \, \mbox{sin}\tilde{\delta} \,L_2^{\;\; 2}$, with
$\mbox{tan} \tilde{\delta} = - k/l$. Then $Spin(7)$ metrics are given by 
\begin{equation} 
	ds_8^2 = dr^2 + a(r)^2 \, \sigma_i^2 + b(r)^2 \, \Sigma_i^2 +
	c(r)^2 \, \nu_i^2  + f^2 \, \lambda^2 , 
\end{equation} 
where the functions $a$, $b$, $c$, $f$  satisfy the differential equations 
\begin{equation} 
	\begin{array}{lcl} 
		\dot{a} &=& \frac{b^2 + c^2 - a^2}{b \, c} -
		\frac{\sqrt{2}\, f\, cos\tilde{\delta}}{a} , \\ 
		\dot{b} &=& \frac{a^2 + c^2 - b^2}{c \, a} + 
		\frac{\sqrt{2}\, f\, sin\tilde{\delta}}{b} , \\  
		\dot{c} &=& \frac{a^2 + b^2 - c^2}{a \, b} + 
		\frac{\sqrt{2}\, f\, (cos\tilde{\delta} -
		sin\tilde{\delta})}{c} , \\ 
		\dot{f} &=&  \frac{\sqrt{2}\, f^2\, (cos\tilde{\delta} -
		sin\tilde{\delta})}{c^2} + \frac{\sqrt{2}\, f^2\,
		cos\tilde{\delta}}{a^2} - \frac{\sqrt{2}\, f^2\,
		sin\tilde{\delta}}{b^2}   .  \\  
	\end{array} 
\end{equation} 
One particular solution is given by 
\begin{equation} 
	\begin{array}{llll} 
		a = A\,  r ,  &   b = B \, r ,  & c = C \, r , & f = F \, r ,
	\end{array} 
\end{equation} 
with 
\begin{equation} 
	\begin{array}{lcl} 
		\frac{l \, F}{\mu \, A^2} = \frac{B^2 + C^2 - A^2}{A\,
		B\, C} - 1 , \hspace{1cm} \mbox{and cyclic} , \\ 
		\left( \frac{l}{A^2} + \frac{k}{B^2} - \frac{l+k}{C^2}
		\right) \, \frac{F}{\mu} = 1 , 
	\end{array} 
\end{equation} 
$\sqrt{2} \, \mu = \sqrt{k^2 + l^2}$. These equations define an almost
$G_2$ metric on the base of the cone.


\newpage 

\begin{thebibliography}{99}

\bibitem{Collins} C.B. Collins, \grg{10}{1979}{925}. 

\bibitem{Yano} K. Yano, \am{55}{1952}{328}. 
 
\bibitem{Floyd} R. Floyd, The dynamics of Kerr fields. Ph.D. thesis,
London University (1973). 
 
\bibitem{Penrose} R. Penrose, Ann. N. Y. Acad. Sci. {\bf 224}, 125
(1973). 

\bibitem{CarterMcLenaghan} B. Carter and R. G. McLenaghan, \prd{19}{1979}{1093}. 
  
\bibitem{GibbonsRietdijkVanHolten} G. W. Gibbons, R. H. Riedtdijk,
J. W. van Holten, \npb{404}{1993}{42-64}. 
 
\bibitem{Tanimoto} M. Tanimoto, \npb{442}{1995}{549-560}. 

\bibitem{Macfarlane_Azcarraga} J. A. de Azc$\acute{a}$rraga,
J. M. Izquierdo, A. J. Macfarlane, \npb{604}{2001}{75-91}. 
 
\bibitem{Macfarlane} A. J. Macfarlane, \npb{621}{2002}{712-722}. 


\bibitem{Kasper1} J.-W. van Holten, K. Peeters and A. Waldron, \cqg{16}{1999}{2537}, 
\hepth{9901163}. 
 
\bibitem{Kasper2} K. Peeters and A. Waldron, \jhep{02}{1999}{24}, 
\hepth{9901016}. 
 

 
\bibitem{BerezinMarinov} F. A. Berezin and M. S. Marinov,
\ap{104}{1977}{336}. 
 
\bibitem{Casalbuoni} R. Casalbuoni, \plb{62}{1976}{49}. 
 
\bibitem{BarducciCasalbuoniLusanna} A. Barducci and  R. Casalbuoni,
L. Lusanna, \nc{35A}{1976}{377}; \npb{124}{1977}{93}. 
 
\bibitem{BrinkDeserZuminoDiVecchiaHowe} L. Brink, S. Deser, B. Zumino,
P. DiVecchia and P. Howe, \plb{64}{1976}{43}. 
 
\bibitem{BrinkDiVecchiaHowe}  L. Brink, P. DiVecchia and P. Howe,
\npb{118}{1977}{76}. 
 
\bibitem{vanHolten} J. W. van Holten,  {\it in}
Proc. Sem. Math. structures in field theories, 1986-87, CWI syllabus
Vol.26 (1990), p.109. 
 
\bibitem{vanHoltenRietdijk} J. W. van Holten and R. H. Rietdijk,
\cqg{7}{1990}{247}. 
 
\bibitem{vanHoltenRietdijk2} J. W. van Holten and R. H. Rietdijk,
\hepth{9205074}; R. H. Rietdijk, Applications of supersymmetric
quantum mechanics, Ph.D. Thesis, Amsterdam (1992). 
 
\bibitem{Riet_VanHolt} R. H. Riedtdijk, J. W. van Holten,
\npb{472}{1996}{427-446}. 
 
\bibitem{Baleanu} D. Baleanu, S. Baskal, \ijmpa{17}{2002}{3737-3748}. 



\bibitem{Tachibana} S. Tachibana, Tohoku Math. J. {\bf 21} (1969) 56. 
 
\bibitem{GlassKress} E. N. Glass and J. Kress, \grqc{9809074}. 
 
\bibitem{FerrandoSaez} J. J. Ferrando and J. A. Saez, \grqc{0212085}. 
 
\bibitem{Stepanov} S. E. Stepanov, \tmp{134(3)}{2003}{333-338}. 

\bibitem{Joyce} D. D. Joyce. Compact Manifolds with Special
Holonomy. Oxford University Press, 2000, pg.66. 

\bibitem{Ortin} N. Alonso-Alberca, E. Lozano-Tellechea and
T. Ort$\acute{i}$n, \hepth{0208158}.  

\bibitem{Bar} C. B$\ddot{a}$r, \cmp{154}{1993}{509-521}.  
 


\bibitem{Gallot} S. Gallot, {\it} Ann. Sci. $\acute{E}$cole Norm. Sup. 
{\bf 12} (1979) 235-267. 

\bibitem{Figueroa} J. M. Figueroa-O'Farrill, \hepth{9902066}. 
 
\bibitem{Friedrich} H. Baum, T. Friedrich, R. Grunewald,
I. Kath. Twistors and Killing spinors on Riemannian manifolds. Sektion
Mathematik der Humboldt-Universit\"{a}t zu Berlin, 1989, 179 S. 
 
\bibitem{Leitner} F. Leitner, \Math{DG}{0302024}.  

\bibitem{Kasper_String} F. De Jonghe, K. Peeters, K. Sfetsos, 
\cqg{14}{1997}{35-46}, \hepth{9607203}. 

 
\bibitem{Boss} M. Cveti$\check{c}$, G. W. Gibbons, H. L\"{u} and
C. N. Pope, \hepth{0108245}. 
 
\bibitem{BilalSfetsos} A. Bilal, J. P. Derendinger and K. Sfetsos, \hepth{0111274}. 

 
 
 
 
 
 

 
 
 


\bibitem{BaleanuBaskalThanksGuys!!!} D. Baleanu and S. Baskal,
\grqc{0206045}. 
 

\bibitem{Witten} E. Witten, \npb{186}{1981}{412}. 

\bibitem{BilalMetzger} A. Bilal and S. Metzger, \hepth{0302021}. 


\bibitem{Sean} S. A. Hartnoll, \plb{532}{2002}{297-304}. 

\bibitem{Boss_BianchiIX}  M. Cveti$\check{c}$, G. W. Gibbons, H. L\"{u} and
C. N. Pope, \hepth{0206151}. 

 
 

 

 




\end{thebibliography}
\end{document}